\documentclass[12pt]{article}
\usepackage{epsfig}
\usepackage{graphicx}
\usepackage{amsmath}
\usepackage{hhline}
\usepackage{amssymb}
\usepackage{times}
\usepackage{cite}

\newlength{\dinwidth}
\newlength{\dinmargin}
\begin{document}  
\newcommand{\pom}{{I\!\!P}}
\newcommand{\reg}{{I\!\!R}}
\newcommand{\slowpi}{\pi_{\mathit{slow}}}
\newcommand{\fiidiii}{F_2^{D(3)}}
\newcommand{\fiidiiiarg}{\fiidiii\,(\beta,\,Q^2,\,x)}
\newcommand{\n}{1.19\pm 0.06 (stat.) \pm0.07 (syst.)}
\newcommand{\nz}{1.30\pm 0.08 (stat.)^{+0.08}_{-0.14} (syst.)}
\newcommand{\fiidiiiful}{F_2^{D(4)}\,(\beta,\,Q^2,\,x,\,t)}
\newcommand{\fiipom}{\tilde F_2^D}
\newcommand{\ALPHA}{1.10\pm0.03 (stat.) \pm0.04 (syst.)}
\newcommand{\ALPHAZ}{1.15\pm0.04 (stat.)^{+0.04}_{-0.07} (syst.)}
\newcommand{\fiipomarg}{\fiipom\,(\beta,\,Q^2)}
\newcommand{\pomflux}{f_{\pom / p}}
\newcommand{\nxpom}{1.19\pm 0.06 (stat.) \pm0.07 (syst.)}
\newcommand {\gapprox}
   {\raisebox{-0.7ex}{$\stackrel {\textstyle>}{\sim}$}}
\newcommand {\lapprox}
   {\raisebox{-0.7ex}{$\stackrel {\textstyle<}{\sim}$}}
\def\gsim{\,\lower.25ex\hbox{$\scriptstyle\sim$}\kern-1.30ex%
\raise 0.55ex\hbox{$\scriptstyle >$}\,}
\def\lsim{\,\lower.25ex\hbox{$\scriptstyle\sim$}\kern-1.30ex%
\raise 0.55ex\hbox{$\scriptstyle <$}\,}
\newcommand{\pomfluxarg}{f_{\pom / p}\,(x_\pom)}
\newcommand{\dsf}{\mbox{$F_2^{D(3)}$}}
\newcommand{\dsfva}{\mbox{$F_2^{D(3)}(\beta,Q^2,x_{I\!\!P})$}}
\newcommand{\dsfvb}{\mbox{$F_2^{D(3)}(\beta,Q^2,x)$}}
\newcommand{\dsfpom}{$F_2^{I\!\!P}$}
\newcommand{\gap}{\stackrel{>}{\sim}}
\newcommand{\lap}{\stackrel{<}{\sim}}
\newcommand{\fem}{$F_2^{em}$}
\newcommand{\tsnmp}{$\tilde{\sigma}_{NC}(e^{\mp})$}
\newcommand{\tsnm}{$\tilde{\sigma}_{NC}(e^-)$}
\newcommand{\tsnp}{$\tilde{\sigma}_{NC}(e^+)$}
\newcommand{\st}{$\star$}
\newcommand{\sst}{$\star \star$}
\newcommand{\ssst}{$\star \star \star$}
\newcommand{\sssst}{$\star \star \star \star$}
\newcommand{\tw}{\theta_W}
\newcommand{\sw}{\sin{\theta_W}}
\newcommand{\cw}{\cos{\theta_W}}
\newcommand{\sww}{\sin^2{\theta_W}}
\newcommand{\cww}{\cos^2{\theta_W}}
\newcommand{\trm}{m_{\perp}}
\newcommand{\trp}{p_{\perp}}
\newcommand{\trmm}{m_{\perp}^2}
\newcommand{\trpp}{p_{\perp}^2}
\newcommand{\alp}{\alpha_s}

\newcommand{\alps}{\alpha_s}
\newcommand{\sqrts}{$\sqrt{s}$}
\newcommand{\LO}{$O(\alpha_s^0)$}
\newcommand{\Oa}{$O(\alpha_s)$}
\newcommand{\Oaa}{$O(\alpha_s^2)$}
\newcommand{\PT}{p_{\perp}}
\newcommand{\JPSI}{J/\psi}
\newcommand{\sh}{\hat{s}}
\newcommand{\uh}{\hat{u}}
\newcommand{\MP}{m_{J/\psi}}
\newcommand{\PO}{I\!\!P}
\newcommand{\xbj}{x}
\newcommand{\xpom}{x_{\PO}}
\newcommand{\ttbs}{\char'134}
\newcommand{\xpomlo}{3\times10^{-4}}  
\newcommand{\xpomup}{0.05}  
\newcommand{\dgr}{^\circ}
\newcommand{\pbarnt}{\,\mbox{{\rm pb$^{-1}$}}}
\newcommand{\gev}{\,\mbox{GeV}}
\newcommand{\WBoson}{\mbox{$W$}}
\newcommand{\fbarn}{\,\mbox{{\rm fb}}}
\newcommand{\fbarnt}{\,\mbox{{\rm fb$^{-1}$}}}
%
%
\newcommand{\qsq}{\ensuremath{Q^2} }
\newcommand{\gevsq}{\ensuremath{\mathrm{GeV}^2} }
\newcommand{\et}{\ensuremath{E_t^*} }
\newcommand{\rap}{\ensuremath{\eta^*} }
\newcommand{\gp}{\ensuremath{\gamma^*}p }
\newcommand{\dsiget}{\ensuremath{{\rm d}\sigma_{ep}/{\rm d}E_t^*} }
\newcommand{\dsigrap}{\ensuremath{{\rm d}\sigma_{ep}/{\rm d}\eta^*} }
\newcommand{\dedx}{\ensuremath{{\rm d} E/{\rm d} x}}
\def\Journal#1#2#3#4{{#1} {\bf #2} (#3) #4}
\def\NCA{Nuovo Cimento}
\def\RPP{Rep. Prog. Phys.}
\def\ARNPS{Ann. Rev. Nucl. Part. Sci.}
\def\NIM{Nucl. Instrum. Methods}
\def\NIMA{{Nucl. Instrum. Methods} {\bf A}}
\def\NPB{{Nucl. Phys.}   {\bf B}}
\def\NPPS{Nucl. Phys. Proc. Suppl.} 
\def\NPPSC{{Nucl. Phys. Proc. Suppl.} {\bf C}}
\def\PR{Phys. Rev.}
\def\PLB{{Phys. Lett.}   {\bf B}}
\def\PRL{Phys. Rev. Lett.}
\def\PRD{{Phys. Rev.}    {\bf D}}
\def\PRC{{Phys. Rev.}    {\bf C}}
\def\ZPC{{Z. Phys.}      {\bf C}}
\def\EJC{{Eur. Phys. J.} {\bf C}}
\def\EPL{{Eur. Phys. Lett.} {\bf}}
\def\CPC{Comp. Phys. Commun.}
\def\NP{{Nucl. Phys.}}
\def\JPG{{J. Phys.} {\bf G}} 
\def\EPC{{Eur. Phys. J.} {\bf C}}
\def\PRSL{{Proc. Roy. Soc.}} {\bf}
\def\PETF{{Pi'sma. Eksp. Teor. Fiz.}} {\bf}
\def\JETPL{{JETP Lett}}{\bf}
\def\IJTP{Int. J. Theor. Phys.}
\def\HJ{Hadronic J.}

 


\begin{titlepage}
\begin{flushleft}
{\tt \today } \\
\end{flushleft}
\vspace*{2.cm}
\begin{center}
\begin{Large}
{\boldmath \bf Simulation of Ultra High Energy Neutrino Interactions in Ice and Water \\}
\vspace*{2.cm} (the ACoRNE Collaboration)$^a$ \\  
\end{Large}
\end{center}
\begin{flushleft}

S.~Bevan$^{1}$,               
S.~Danaher$^{2}$,               
J.~Perkin$^{3}$,                
S.~Ralph$^{3 \dagger}$,                   
C.~Rhodes$^{4}$,                    
L.~Thompson$^{3}$,                  
T.~Sloan$^{5 b}$              
and
D.~Waters$^{1}$.                 

\bigskip{\it

 $ ^{1}$ Physics and Astronomy Dept, University College London, UK.  \\

 $ ^{2}$ School of Computing Engineering and Information Sciences, University of Northumbria, Newcastle-upon-Tyne, UK.  \\

 $ ^{3}$ Dept of Physics and Astronomy, University of Sheffield, UK.  \\

 $ ^{4}$ Institute for Mathematical Sciences, Imperial College London, UK.\\

 $ ^{5}$ Department of Physics, University of Lancaster,
          Lancaster, UK \\

\bigskip

\smallskip
 $ ^{\dagger}$ Deceased \\

\bigskip
$ ^a$ Work supported by the UK Particle Physics and Astronomy Research
      Council and by the Ministry of Defence Joint Grants Scheme \\
$^b$ Author for correspondence, email t.sloan@lancaster.ac.uk \\
}

\end{flushleft}

\vspace*{2.5cm}

\begin{abstract}
\noindent
The CORSIKA program, usually used to simulate extensive cosmic ray 
air showers, has been adapted to work in a water or ice medium. 
The adapted CORSIKA code was used to simulate hadronic showers 
produced by neutrino interactions. The simulated showers  
have been used to study the spatial distribution of the 
deposited energy in the showers. This allows a more precise 
determination of the acoustic signals produced by ultra high energy 
neutrinos than has been possible previously. The properties of 
the acoustic signals generated by such showers are described.  

\end{abstract}

\begin{center}
(Submitted to Astroparticle Physics)
\end{center}

\end{titlepage}


\newpage
\section{Introduction}
\label{Intro}

In recent years interest has grown in the detection of very high energy 
cosmic ray neutrinos \cite{ARENAs}. Such particles could be produced 
in the cosmic particle accelerators which make the charged primaries 
or they could be produced by the interactions of the primaries with 
the Cosmic Microwave Background, the so called GZK effect \cite{GZK}. 
The flux of neutrinos expected from these two sources has been 
calculated \cite{WBlim,GZKnu}. It is  found to be very low so 
that large targets are needed for a measurable detection rate.  
It is interesting to measure 
this neutrino flux to see if it is compatible with the values 
expected from these sources, incompatibility implying   
new physics.   

Searches for cosmic ray neutrinos are ongoing in AMANDA \cite{AMANDA}, 
IceCube \cite{IceCube}, ANTARES \cite{Antares} and NESTOR \cite{NESTOR}, 
detecting upward going 
muons from the Cherenkov light in either ice or water.  
In general, these experiments are sensitive to lower energies than 
discussed here since the Earth becomes opaque to neutrinos 
at very high energies. The experiments could detect  
almost horizontal higher energy neutrinos but have limited target 
volume due to the attenuation of the light signal in the ice. 
The Pierre Auger Observatory, an extended air shower array detector,  
will also search for upward and almost horizontal showers from 
neutrino interactions \cite{Auger}.   
In addition to these detectors there are ongoing experiments to detect 
the neutrino interactions by either radio or acoustic emissions from 
the resulting particle showers \cite{ARENAs}. 
These latter techniques, with much longer attenuation lengths, 
allow very large target volumes utilising either large ice fields 
or dry salt domes for radio or ice fields, salt domes and the oceans 
for the acoustic technique. 

In order to assess the feasibility of each technique the production of 
the particle shower from neutrino interactions needs to be simulated.
Since experimental data on the interactions of such high energy particles 
do not exist it is necessary to use theoretical models to simulate them. 
The most extensive ultra high energy simulation program which has so far 
been developed is CORSIKA \cite{CORSIKA}. However, this program has been 
used previously only for the simulation of cosmic ray air showers.  
The program is readily available \cite{CORSIKA}.  

Different simulations are necessary for the radio and acoustic techniques. 
Radio emission occurs due to coherent Cherenkov 
radiation from the particles in the shower, the Askaryan Effect \cite{Asky}. 
The emitted energy is sensitive to the distribution of the 
electron-positron asymmetry which develops in the shower and which grows 
for lower energy electromagnetic particles. Hence, to simulate radio 
emission, the 
electromagnetic component of the shower must be followed down to 
very low kinetic energies ($\sim 100$ keV)\cite{JZ}. In contrast, 
an acoustic signal is generated by the sudden local heating of the 
surrounding medium induced by the particle shower \cite{Learned}. 
Thus to simulate the acoustic signal the spatial distribution of 
the deposited energy is needed. Once the electromagnetic energy in 
the shower reaches the MeV level (electron range $\sim 1$ cm) the 
energy can be simply added to the total deposited energy  
and the simulation of such particles discontinued. 
Extensive simulations have been carried 
out for the radio technique \cite{Zasetal}. However, the simulations for 
the acoustic technique are less advanced. Some work has been done 
\cite{BN,Valentin} 
using the Geant4 package \cite{Geant4}. However, this work is 
restricted to energies less than $10^5$ GeV for hadron showers since 
the range of validity of the physics models in this package does not 
extend to higher energy hadrons.  

In this paper the energy distributions of showers produced by 
neutrino interactions in sea water at energies up to $10^{12}$ GeV 
are discussed. The distributions are generated using the  
air shower program CORSIKA \cite{CORSIKA} modified to work in a sea 
water medium. The salt component of the sea water has a negligible 
effect\footnote{The shower maximum was observed to peak at a depth 
$2.4 \pm 1.1\%$ less in sea water than in fresh water with the same peak 
energy deposited, for protons of energy $10^5$ GeV.}
 and the results are presented in distance units of g cm$^{-2}$, 
hence they should be applicable to ice also. 
The computed distributions have been parameterised and this 
parameterisation is used to develop a simple program 
to simulate neutrino interactions and the resulting particle showers. 
The properties of the acoustic signals from the generated showers 
are also presented.  

\section{Adaptation of the CORSIKA program to a water medium}

The air shower program, CORSIKA (version 6204) \cite{CORSIKA}, has 
been adapted to run 
in sea water i.e. a medium of constant density of 1.025 g cm$^{-3}$ 
rather than the variable density needed for an air atmosphere. Sea 
water was assumed to consist of a medium in which $66.2\%$ of the 
atoms are hydrogen, $33.1\%$ of the atoms are oxygen and $0.7\%$ of the 
atoms are made of common salt, NaCl. The salt was assumed to be 
a material with atomic weight and atomic number A=29.2 and Z=14, the mean 
of sodium and chlorine. The purpose of this is to maintain the structure 
of the program as closely as possible to the air shower version 
which had two principal atmospheric components (oxygen and nitrogen) 
with a trace of argon. The presence of the salt component had an
almost undetectable effect on the behaviour of the showers. 

Other changes made to the program to accommodate the water medium 
include a modification of the stopping power formula to allow for 
the density effect in water
\footnote{The stopping power was computed using the Bethe-Bloch 
formula \cite{PDG} and the density effect from the formulae of 
Sternheimer et al \cite{Rudi}.}. This only affects the energy loss 
for hadrons since the stopping powers for electrons are part of the 
EGS \cite{EGS} package which is used by CORSIKA to simulate the 
propagation of the electromagnetic component of the shower. 
Smaller radial binning of the shower was also required since shower 
radii in water are much smaller than those in air. In addition the 
initial state energy for electrons and photons  
above which the LPM effect \cite{LPM} was simulated 
in the program was reduced to the much lower value necessary for 
water \footnote{The level was set at 1 TeV compared to the 
characteristic energy for water $E_{LPM}=270$ TeV \cite{PDG}.}. 
The LPM effect suppresses pair production from high energy photons 
and bremsstrahlung from high energy electrons.  
Similarly, the interactions of neutral pions had to be simulated at 
lower energy than in air because of the higher density water medium.  
In all about 100 detailed changes needed to be made to the CORSIKA 
program to accommodate the water medium.

\begin{figure}[h]
\includegraphics[width=36pc]{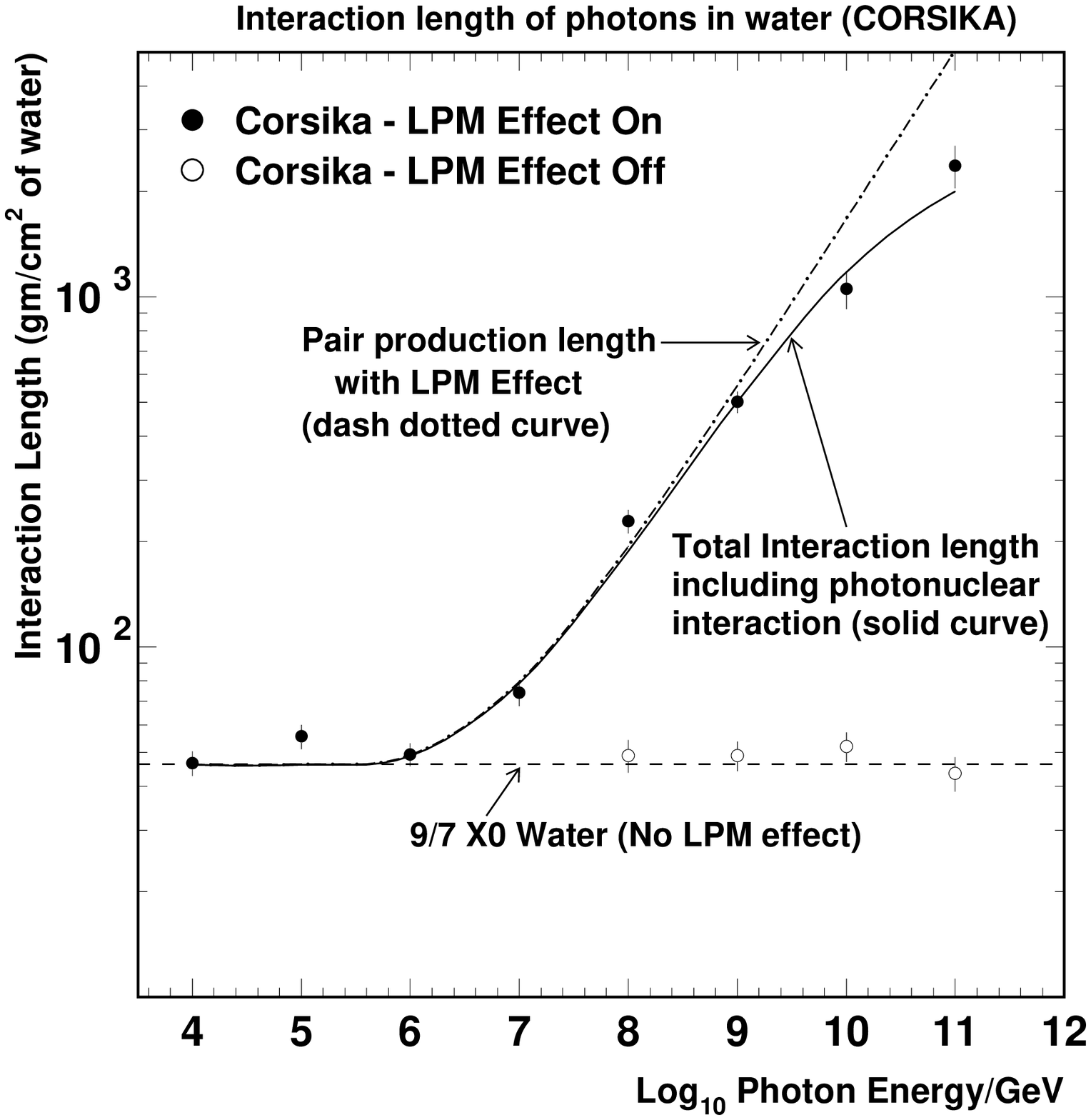}\hspace{2pc}%
\caption{\label{LPMfig}The interaction length for high energy gamma 
rays versus the photon energy measured in CORSIKA (data points with 
statistical errors). The dash dotted curve shows the pair production 
length computed from the LPM effect using the formulae of 
Migdal \cite{LPM}. The solid curve shows the computed total interaction 
length, including both pair production and photonuclear interactions 
with the cross section from CORSIKA. The dashed line labelled $9/7 X_0$ 
shows the expected pair production length without the LPM effect. Here 
$X_0$ is the radiation length of the material.}
\end{figure}

To test the implementation of the LPM effect \cite{LPM} in the program 
100 showers from incident gamma ray photons at several different energies 
were generated and the mean 
depth of the first interaction (the mean free path) calculated.  
The observed mean free path was found to be in agreement with the 
expected behaviour when both the suppression of pair production 
and photonuclear interactions were taken into account 
(see Figure \ref{LPMfig}). This showed that 
the LPM effect had been properly implemented in CORSIKA.  

Considerable fluctuations between showers occurred. These are expressed 
in terms of the ratio of the root mean square (RMS) deviation of a given 
parameter to its mean value: the RMS peak energy deposit to  
the mean peak energy deposit was observed to be $14\%$ at $10^5$ GeV reducing 
to $4\%$ at $10^{11}$ GeV, that for the depth of the peak position varied   
from $19\%$ to $7.4\%$ and for the full width at half maximum 
of the shower from $63\%$ to $18\%$.
To smooth out such fluctuations averages of 100 generated showers will 
be taken in the following. The statistical error on the averages is then 
given by these RMS  values divided by 10. The hadronic energy 
contributes only about $10 \%$ to the shower energy at the shower peak, 
the remainder being carried by the electromagnetic part of the shower. 
 
The simulations were all carried out in a vertical column of sea water 
20 m long. The deposited energy generated by CORSIKA was binned into 
20 g cm$^{-2}$ slices longitudinally and 1.025 g cm$^{-2}$ annular 
cylinders radially for $0 < r < 10.25$ g cm$^{-2}$ and 10.25 g cm$^{-2}$ for 
$10.25 < r < 112.75$ g cm$^{-2}$ where $r$ is the distance from the 
vertical axis. To reduce computing times, the thinning option was 
used i.e. below a certain fraction of the 
primary energy (in this case $10^{-4}$) only one of the particles 
emerging from the interaction is followed and an appropriated weight 
is given to it \cite{CORSIKA_PHYSICS.ps.gz}. The simulation of particles 
continued down to cut-off energies of 3 MeV for electromagnetic particles 
and 0.3 GeV for hadrons. When a particle reached this cut-off, the energy 
was added to the slice where this occurred.  The 
QGSJET \cite{QGSJET} model was used to simulate the hadronic interactions.

\section{Comparison with other simulations}

\subsection{Comparison with Geant4}

Proton showers were generated in sea water using the program 
Geant4 (version 8.0) \cite{Geant4} and compared with those 
generated in CORSIKA. 
Unfortunately, the range of validity of Geant4 physics models for hadronic 
interactions does not extend beyond an energy of $10^5$ GeV. Hence the 
comparison is restricted to energies below this.  

\begin{figure}[h]
\includegraphics[width=36pc]{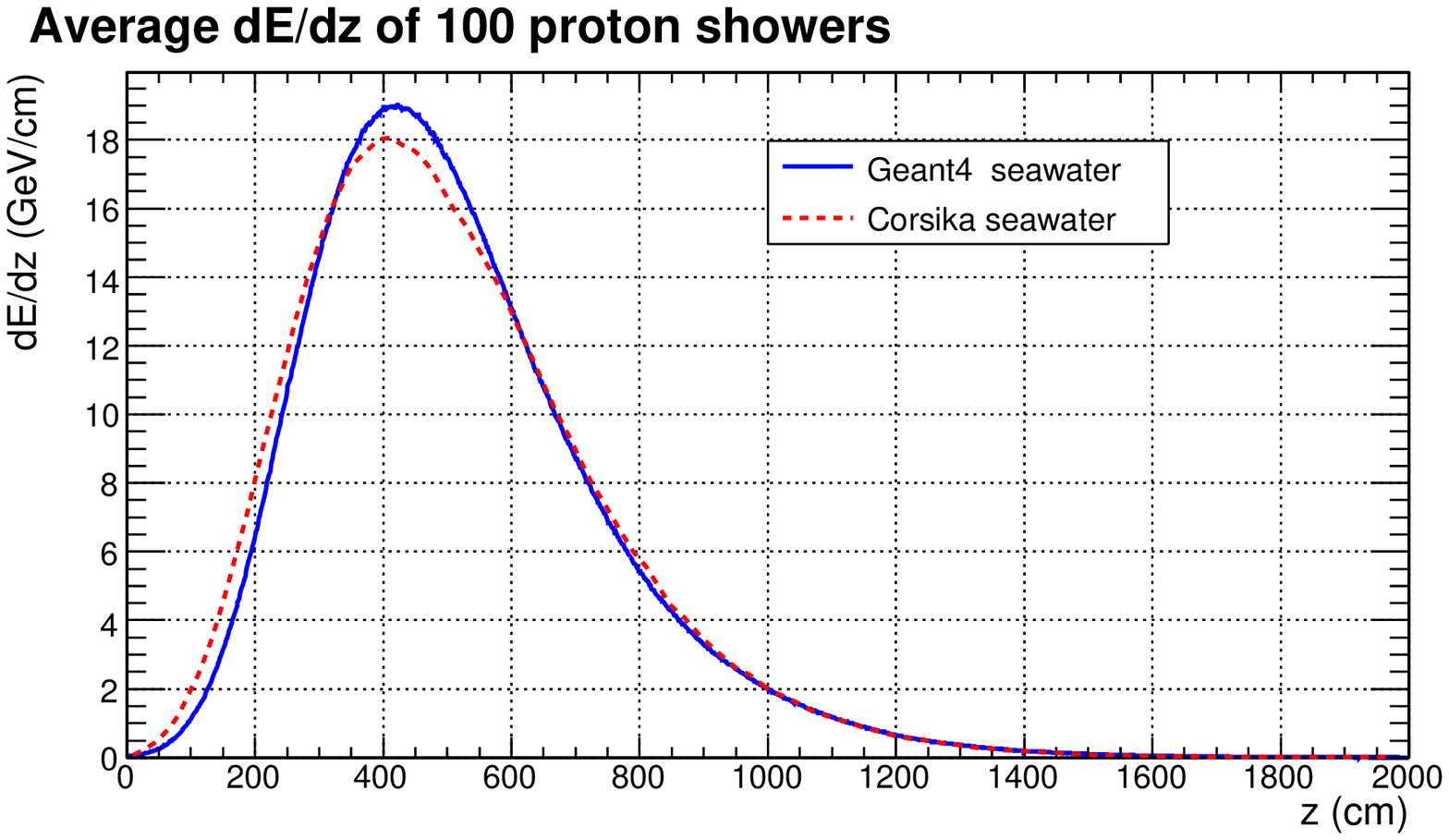}\hspace{2pc}%
\includegraphics[width=36pc]{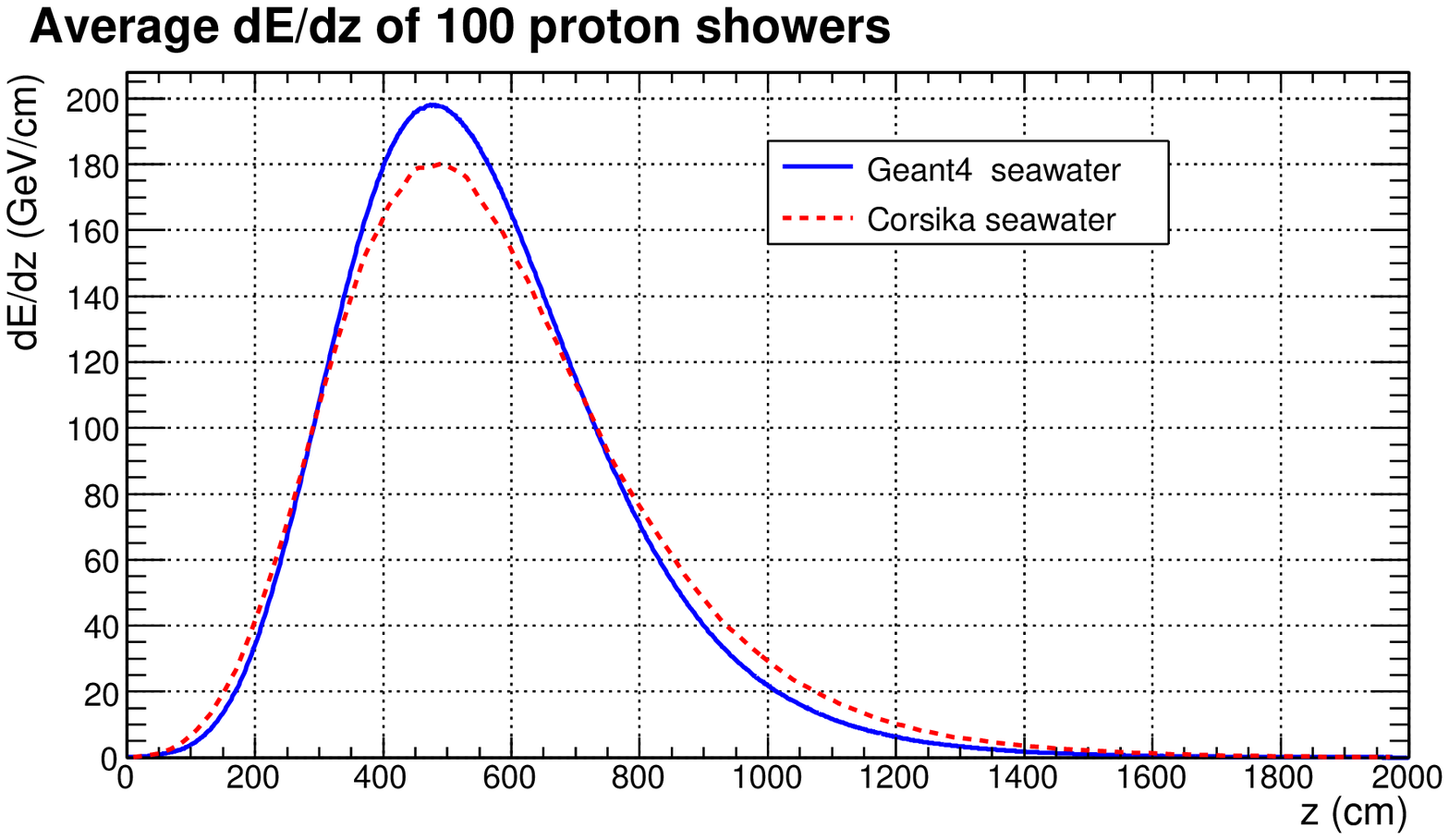}\hspace{2pc}%
\caption{\label{long}Averaged longitudinal energy deposited per unit 
path length of 100 proton showers at energy $10^4$ GeV (upper plot) 
and $10^5$ GeV (lower plot) generated in 
Geant4 and CORSIKA versus depth in the water. }
\end{figure}

\begin{figure}
\begin{minipage}{20pc}
\includegraphics[height=50pc,width=21pc]{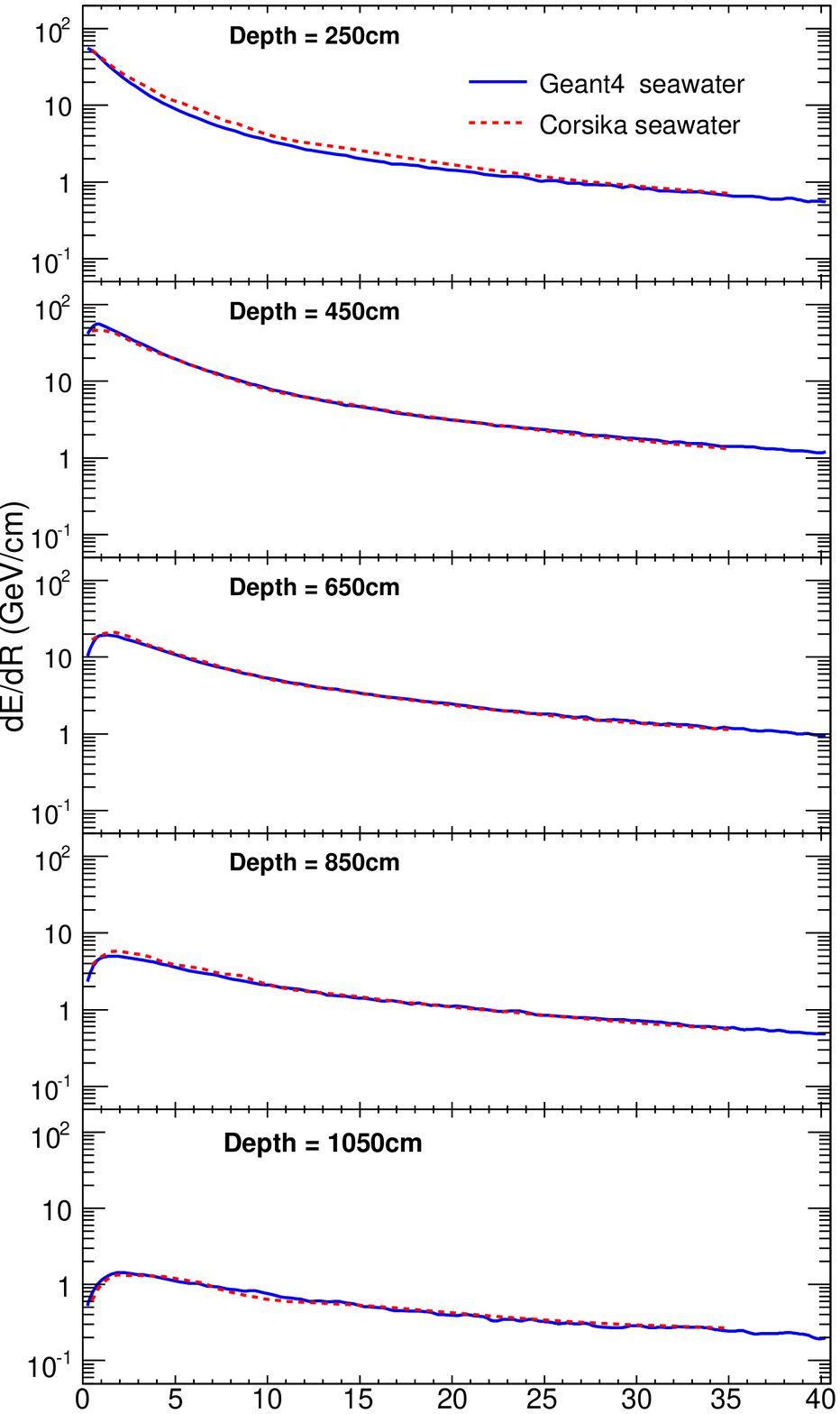}
\end{minipage}\hspace{-1pc}%
\begin{minipage}{20pc}
\includegraphics[height=50pc,width=21pc]{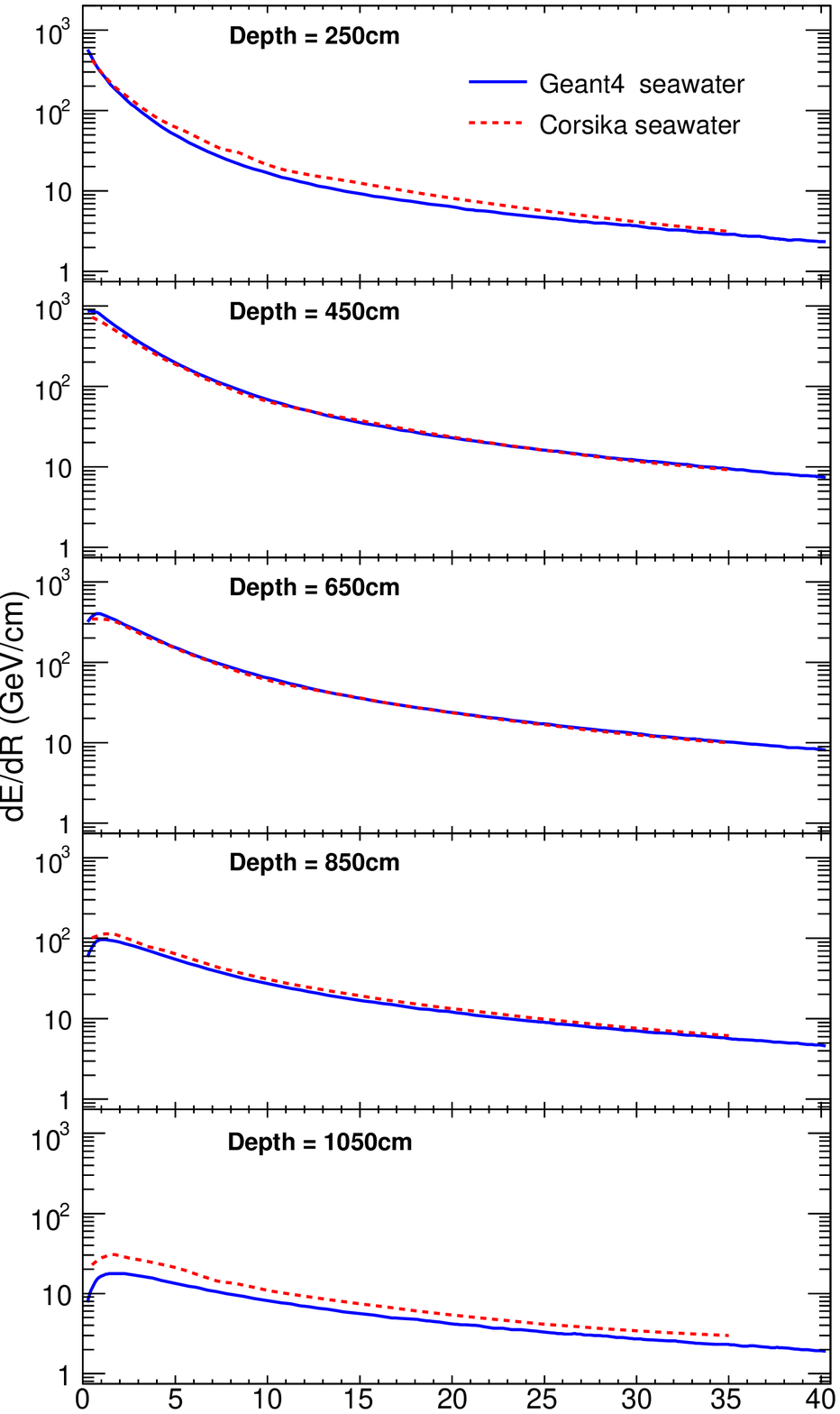}
\end{minipage}\hspace{-1pc}
\vspace*{-5mm}
\caption{\label{rad}Averaged radial energy deposited per 20 g cm$^{-2}$ 
vertical slice per unit radial distance for 100 proton showers at 
energy $10^4$ GeV (left hand plots) and $10^5$ GeV (right hand plots) 
generated in Geant4 and CORSIKA versus distance from 
the axis in the water for different depths of the shower.}
\end{figure}

Figure \ref{long} shows the longitudinal distributions of proton showers 
at energies of $10^4$ and $10^5$ GeV (averaged over 100 showers) 
as determined from Geant4 and CORSIKA. The showers from CORSIKA tend to 
be slightly broader and with a smaller peak energy than those generated 
by Geant4. The difference in the peak height is $\sim 5\%$ at $10^4$ GeV 
rising to $\sim 10\%$ at energy $10^5$ GeV. Figure \ref{rad} shows the 
radial distributions. The differences in the longitudinal distributions 
are reflected in the radial distributions. However, the shapes of the 
radial distributions are very similar between Geant4 and CORSIKA, with  
CORSIKA producing $\sim10\%$ more energy near 
the shower axis at depths between 450 and 850 g cm$^{-2}$ where most 
of the energy is deposited. The acoustic signal from a shower is most 
sensitive to the radial distribution, particularly  near the 
axis ($r \sim 0$). It is relatively insensitive to the shape of the 
longitudinal distribution.     

\subsection{Comparison with the simulation of  Alvarez-Mu\~niz 
and Zas}

The CORSIKA simulation was also compared with the longitudinal shower 
profile for protons computed in the simulation by Alvarez-Mu\~niz and 
Zas (AZ) \cite{AZpaper}.
There was a reasonable agreement between the longitudinal shower shapes 
from CORSIKA and those shown in Figure 2 of ref.~\cite{AZpaper}. However, 
the numbers of electrons and positrons at the peak of the CORSIKA showers 
was $\sim 20\%$ lower than those from ref.~\cite{AZpaper}. 
This number is sensitive to the energy below which these 
particles are counted and this is not specified in \cite{AZpaper}. 
Hence the agreement between CORSIKA and their simulation is probably 
satisfactory within this uncertainty.      
 

In conclusion, the modifications made to CORSIKA to simulate high energy 
showers in a water medium give results which are compatible with the 
predictions from the Geant4 simulations for energy less than $10^5$ GeV 
and the simulation of AZ within $20\%$. This is 
taken to be the accuracy of the simulation program assuming that there 
are no unexpected and unknown interactions between the centre of mass 
energy explored at current accelerators and those studied in 
these simulations. 
Studies of the sensitivity of the CORSIKA simulation to the different 
models of the hadronic interactions have been reported in 
reference \cite{Knapp}. They find that the peak number of electrons plus 
positrons varies by $\sim 20\%$  for proton showers in air depending on 
the choice of the hadron interaction model used. These differences are 
similar in magnitude to the differences between the AZ, Geant4 and CORSIKA 
simulations reported here. Hence the observed 
differences between the Geant4, AZ and CORSIKA simulations in water 
could be within the uncertainties of the hadronic interaction models. 

\section{Simulation of neutrino induced showers}

Neutrinos interact with the nuclei of the detection medium by either 
the exchange of a charged vector boson ($W^+$), i.e. charged current (CC) 
interactions or the exchange of the neutral vector boson ($Z^0$), 
i.e. neutral current (NC) deep inelastic scattering interactions 
(see for example \cite{Renton}). The ratio of the CC to NC interaction 
cross sections is approximately 2:1. The CC interactions produce 
charged secondary scattered leptons while the NC interactions produce  
neutrinos. The hadron shower carries a fraction $y$ of the energy of 
the incident neutrino and the scattered lepton the remaining fraction 
$1-y$.  We assume that the neutrino flavours are homogeneously mixed 
when they arrive at the Earth by neutrino oscillations. Hence in the CC 
interactions electrons, $\mu$ and $\tau$ leptons will be produced as the 
scattered leptons in equal proportions. At the energies we shall consider,  
these particles behave in a manner similar to minimum ionising particles 
for $\mu$ and $\tau$ leptons. This is almost true also for electrons 
for which the bremsstrahlung process will be suppressed by the 
LPM effect. Hence the charged scattered leptons contribute little    
to the energy producing an acoustic signal. In the 
case of NC interactions there is no contribution to this  energy from 
the scattered lepton. For these reasons the contribution of the scattered 
lepton to the shower profile is ignored beyond $z=20$ m in what follows. 

It is interesting to note that a $\tau$ lepton can decay to hadrons or 
a very high energy electron or muon can produce bremsstrahlung photons at 
large distances from the interaction point. These can initiate further 
distant showers, the so called ``double bang'' effect. The stochastic 
nature of such electron showers is studied in \cite{BN,Valentin}. 
These effects are not considered in this study.

\subsection{Neutrino-nucleon interaction cross sections.}
A number of groups have computed the high energy neutrino-nucleon  
interaction cross sections, $\sigma$, \cite{Ghandi,mks,GYY}. 
In the quark parton model of the nucleon for the single vector boson 
exchange process, the differential cross section for 
CC interactions can be expressed in terms of the measured 
structure functions of the target nucleon $F_2$ and $xF_3$ as 
\begin{equation}
\frac{d^2 \sigma}{dQ^2 dy}= \frac{G_F^2}{2 \pi y}\bigg(\frac{M_W^2}{Q^2+M_W^2}
\bigg)^2 (F_2(x,Q^2)(1-y+y^2/2) \pm y(1-y/2)xF_3(x,Q^2))
\label{sigSF}
\end{equation}
where $G_F$ is the Fermi weak coupling, $M_W$ is the mass of the 
weak vector boson, $Q^2$ is the square of the four momentum transferred to 
the target nucleon, $y=\nu/E$ where $\nu$ is the energy transferred to 
the nucleon ($\nu=E-E^\prime$ with $E$ and $E^\prime$ the energies of the 
incident and scattered leptons) and $x=Q^2/2M\nu$ is the fraction of the 
momentum of the target nucleon carried by the struck quark (here $x$ and $y$ 
are defined for a stationary target nucleon). 
The plus (minus) sign is for neutrino (anti-neutrino) interactions. It can 
be seen that $y$ is the fraction of the neutrino's energy which is converted   
into the energy of the hadron shower. A similar expression can be 
written down for the NC interaction (see for example 
\cite{Renton}) which has a ratio to the CC cross section 
varying from 0.33 to 0.41 as the neutrino energy increases from $10^4$ to 
$10^{13}$ GeV. The structure functions $F_2$ and $xF_3$ are the sum of 
the quark distribution functions which have been parameterised by fitting 
data \cite{MRS99,cteq}. It can be shown that $Q^2=sxy$ where $s=2ME$ is 
the square of the centre of mass energy ($M$ is the target nucleon mass). 
To compute the cross sections the structure functions must be calculated at 
values of $x \lesssim M_W^2/s$ 
i.e. at values well outside the region of the fits to the 
parton distribution functions (PDFs) which have been performed 
for $x \gtrsim 10^{-5}$, 
the range of current measurements. The extrapolation outside the measurement 
range is discussed in \cite{mks}, \cite{Ghandi} and \cite{KMS1,KMS2}. 
Here we adopt the procedure of extrapolating linearly on a log-log scale 
from the parameterised parton distribution functions of \cite{MRS99} 
computed at $x=10^{-4}$ and $x=10^{-5}$. By considering various theoretical 
evolution procedures it is estimated in \cite{Ghandi} that the procedure 
has an accuracy of $\sim 32\%$ per decade and we use this as an 
estimate of the accuracy of the calculation. However, this could be an 
underestimate \cite{RST}.  

\begin{figure}[h]
\includegraphics[width=36pc]{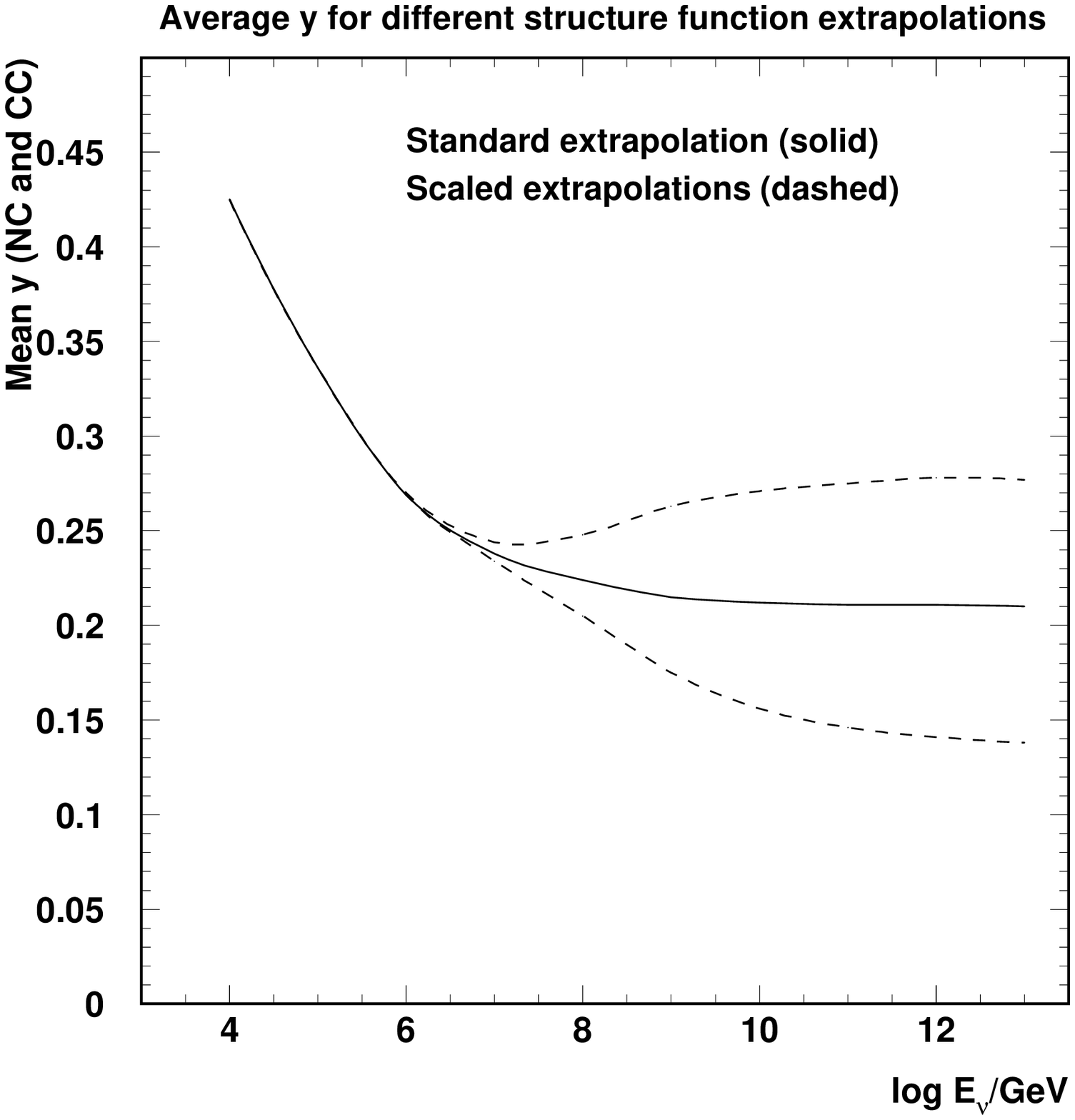}\hspace{2pc}%
\caption{\label{yave}
The mean value of $y$ as a function of energy for $\nu_\mu$ 
interactions  computed according 
to the standard model with the PDFs of MRS99    
\cite{MRS99}, extrapolating $x$ and $Q^2$ out of the fit range from  
$x = 10^{-4}$  linearly on a log-log scale. The upper dashed curve 
shows the result of multiplying the PDFs by $1.32^{\log (10^{-4}/x)}$ for 
PDFs with $x < 10^{-4}$ and the lower dashed curve by dividing by this 
factor. The deviations of the dashed curve from the solid one is an 
indication of the precision of the standard model.}
\end{figure}

\begin{figure}[h]
\includegraphics[width=36pc]{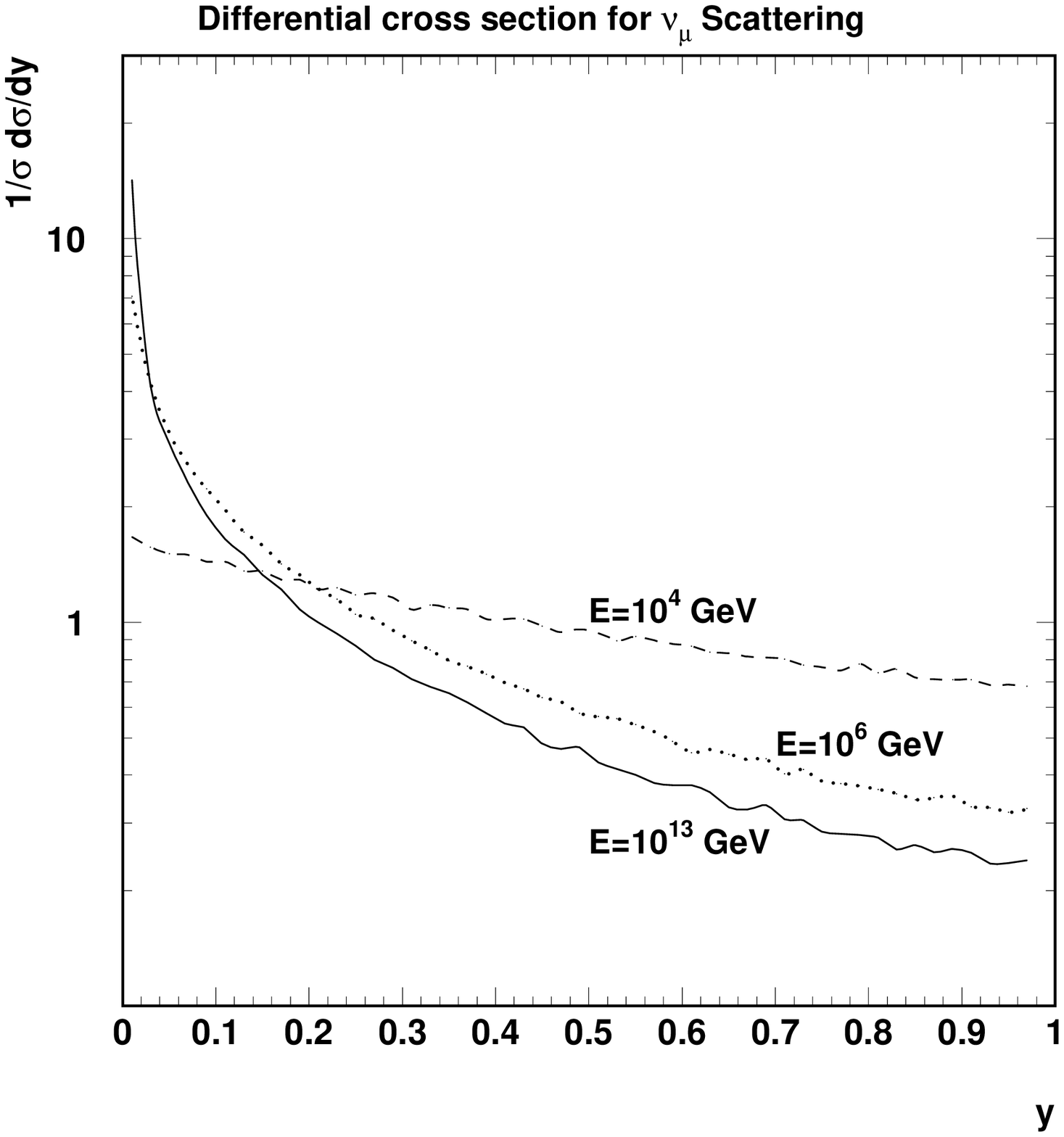}\hspace{2pc}%
\caption{\label{dsdy}
The fraction of events per unit $y$ interval for different $\nu_\mu$ 
energies computed by integrating the expressions for the CC and NC 
cross sections.}
\end{figure}

The expression in equation 
\ref{sigSF} for charged current interactions and the one for neutral current 
interactions were integrated to obtain the total neutrino-nucleon 
interaction cross section, the value of 
the fraction of events per interval of $y$, $1/\sigma d\sigma/dy$, and 
the mean value of $y$. The total cross section was found to be in good 
agreement with the values in \cite{mks,Ghandi} and in reasonable 
agreement with \cite{GYY} which is based on a model different from the 
quark parton model.  Figure \ref{yave} shows the mean value of 
$y$ obtained from this procedure (solid curve) and the effect of 
multiplying or dividing the PDFs by a factor 1.32 per decade (dashed 
curves) as an indication of the possible range of uncertainties in 
the extrapolation of the PDFs. Figure \ref{dsdy} shows the $y$ dependence 
of the cross section for different neutrino energies.   
  
\subsection{A simple generator for neutrino interactions.}

A simple generator for neutrino interactions in a column of water of 
thickness 20 m  was constructed as follows. The neutrino interacts at the 
top of the water column (z=0, with the z axis along the axis of the column).  
The energy fraction transferred, $y$, for the interaction was generated, 
distributed according to 
the curve for the energy of the neutrino shown in Figure \ref{dsdy}. 
This allows the energy of the hadron shower to be calculated for the event. 
The assumption was made that these hadron showers will have approximately 
the same distributions as those of a proton interaction at z=0 (see 
Section \ref{Hernu} for a test of this assumption).   
A series of files of 100 such proton interactions were generated at 
energies in steps of half an order of magnitude between $10^5$ and 
$10^{12}$ GeV. The hadron shower for each neutrino interaction was 
selected at random from the 100 showers in the file at the proton 
energy closest to the energy of the hadron shower. The deposited energy 
in each bin was then multiplied by the ratio of the energy of the 
hadron shower to that of the proton shower. This is made possible 
because the shower shapes vary slowly with shower 
energy. For example, the ratio of the peak energy deposit per 20 
g cm$^{-2}$ slice to the shower energy varies from 0.037 to 0.030 
as the proton shower energy varies from  $10^5$ to $10^{12}$ GeV.   


\subsection{The HERWIG neutrino generator.}
\label{Hernu}

The CORSIKA program has an option to simulate the interactions 
of neutrinos at a fixed point \cite{Ofelia}. The first interaction 
is generated by the HERWIG package \cite{Herwig}. This option 
was adapted to our version of CORSIKA in sea water.  
Some problems were encountered with the $y$ dependence of the 
resulting interactions due to the extrapolation of the PDFs 
to very small $x$ at high energies. This only affects the rate 
of the production of the showers at different $y$ and  
the distribution of the hadrons produced in the interaction 
at a given $y$ should be unaffected. 

\begin{figure}[h]
\includegraphics[width=36pc]{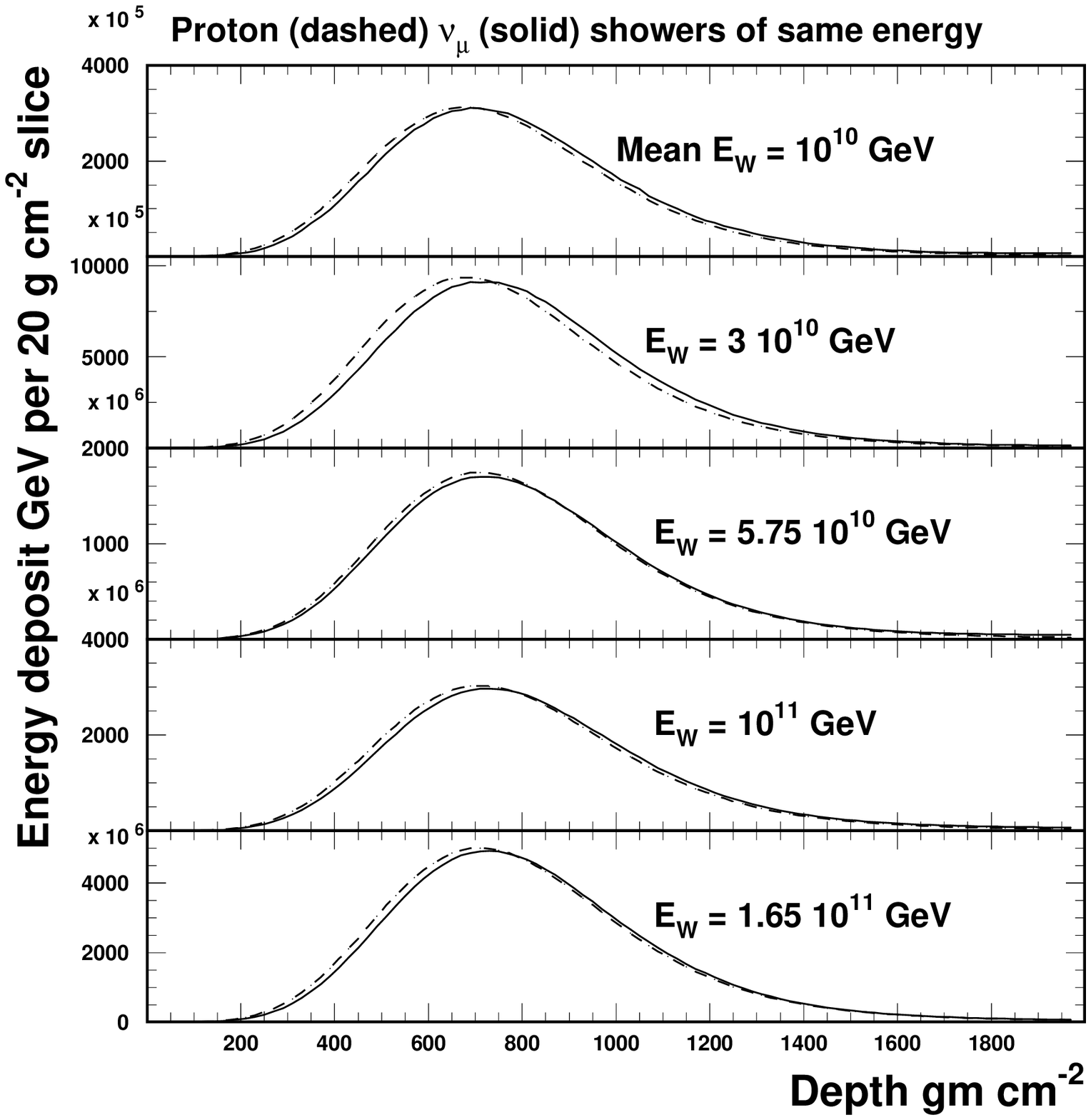}\hspace{2pc}%
\caption{\label{pnul}
The longitudinal distribution of the deposited energy for neutrino 
showers (solid) generated by the Herwig-CORSIKA package and proton 
showers (dashed) scaled to the same values of shower energy $E_W$. The scaling 
factors applied to the average of the protons showers with energy 
$10^{10}$ GeV were 1.0 and 3.0 for  $E_W= 10^{10}$ GeV and 
$E_W=3 \cdot 10^{10}$ GeV, respectively. Those applied to proton showers 
with energy $10^{11}$ GeV were 0.575, 1.0 and 1.65 for 
$E_W=5.75 \cdot 10^{10}$ 
$E_W=10^{11}$ and $E_W=1.65 \cdot 10^{11}$ GeV, respectively.}
\end{figure}

\begin{figure}
\begin{minipage}{20pc}
\includegraphics[height=50pc,width=21pc]{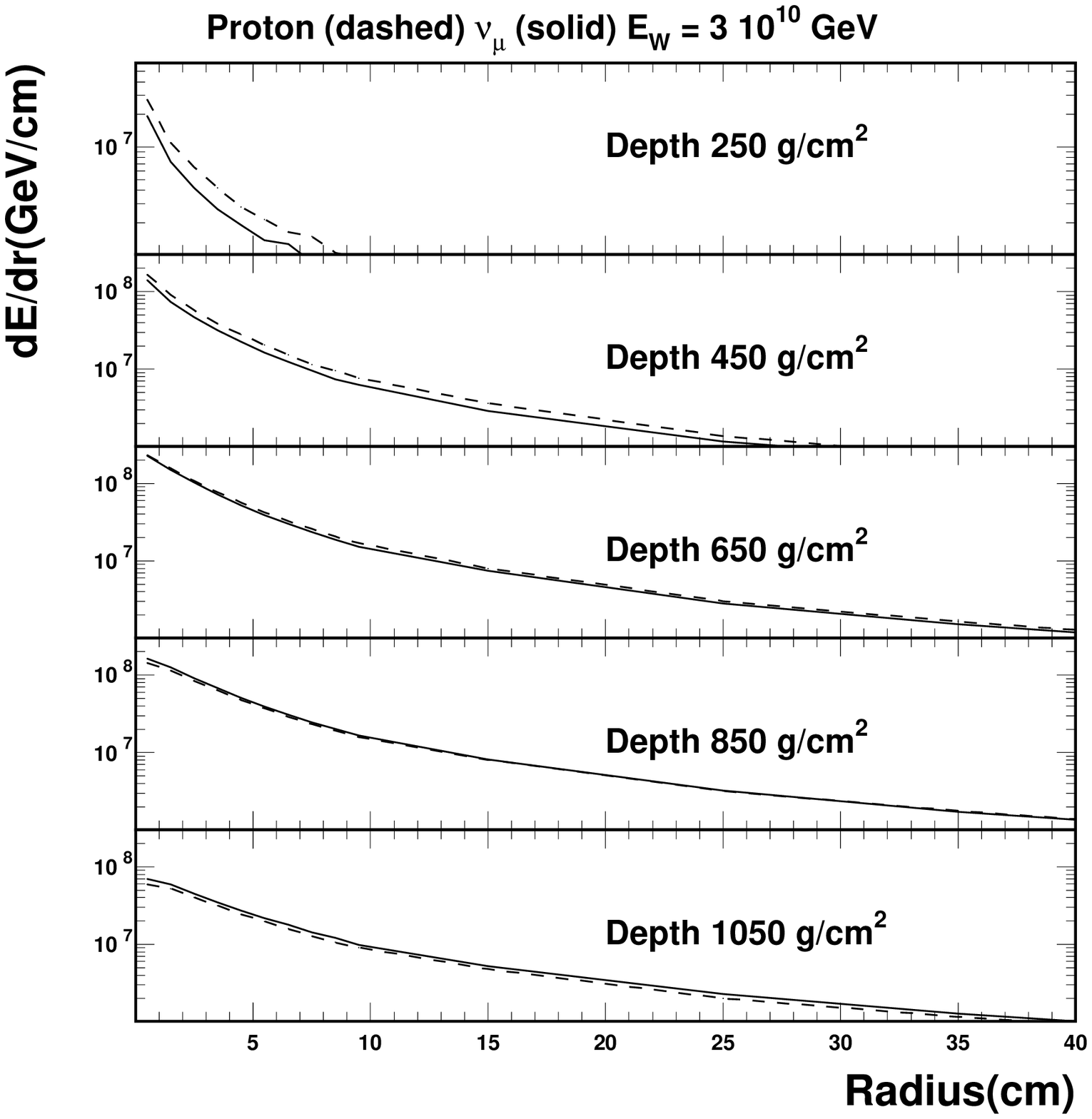}
\end{minipage}\hspace{-1pc}%
\begin{minipage}{20pc}
\includegraphics[height=50pc,width=21pc]{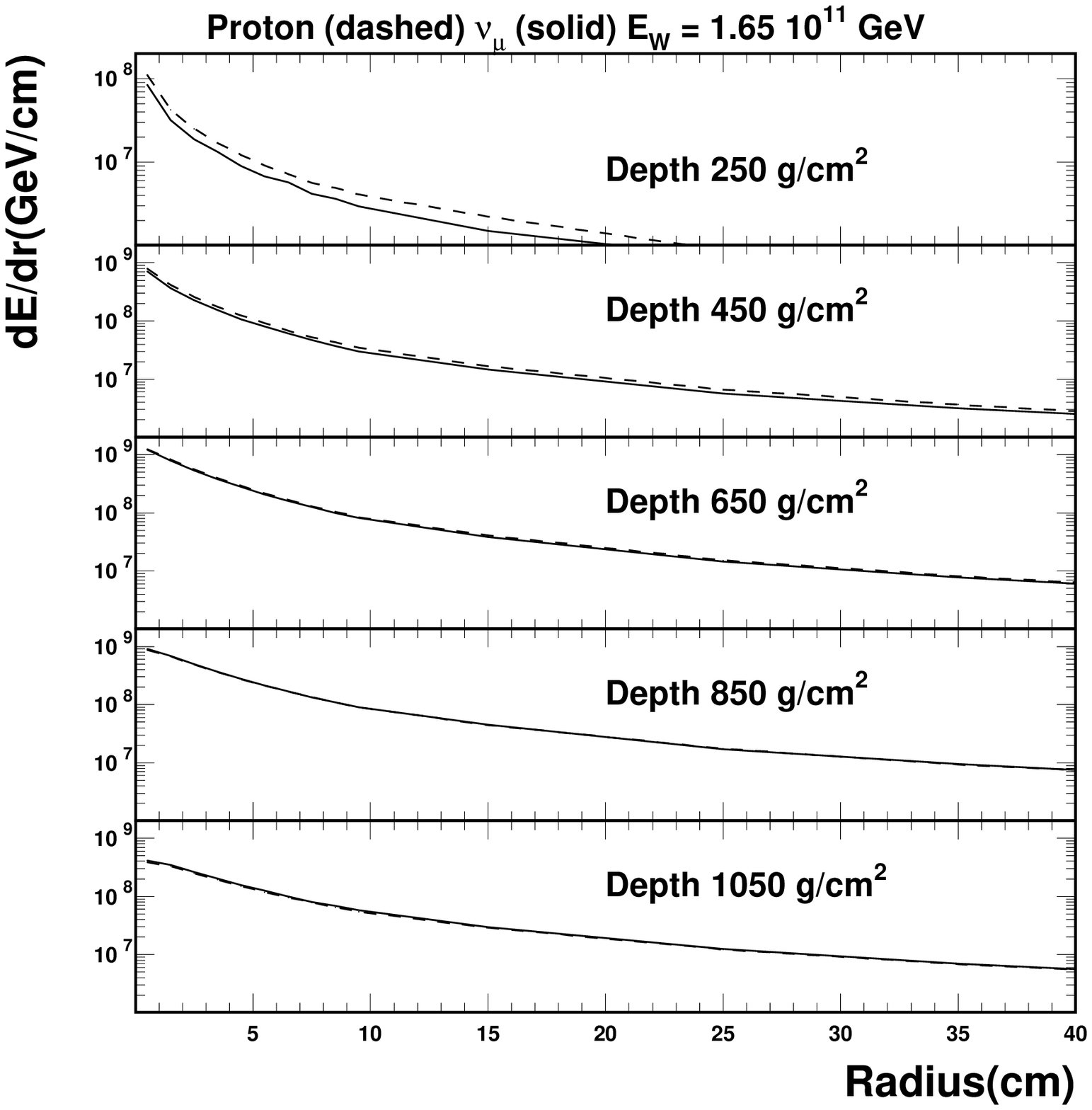}
\end{minipage}\hspace{-1pc}
\vspace*{-5mm}
\caption{\label{pnutr}The solid curves show the averaged radial energy 
deposited per 20 g cm$^{-2}$ 
vertical slice per unit radial distance for 70 neutrino showers with energy 
transfer $E_W = 3 \cdot 10^{10}$ GeV (left hand plots) 
and $E_W = 1.65 \cdot 10^{11}$ GeV (right hand 
plots). The incident neutrino energy was $2 \cdot 10^{11}$ GeV. For comparison 
the dashed curves show the distributions from proton showers scaled to 
these energies. In the left (right) hand plots protons of energy $10^{10}$ 
GeV ($10^{11}$ GeV) were scaled by a factor of 3 (1.65).}
\end{figure}

A total of 700 neutrino 
interactions were generated at an incident neutrino energy of 
$2 \cdot 10^{11}$ GeV. These were divided into the shower energy 
intervals $0.5-2 \cdot 10^{10}$, $2-4 \cdot 10^{10}$, $4-7.5 \cdot 10^{10}$, 
$0.75-1.3 \cdot 10^{11}$ and $1.3-2 \cdot 10^{11}$. The showers in which the 
scattered lepton energy disagreed with the shower energy by more 
than $20\%$ were eliminated leading to a loss of $17\%$ of the 
events with shower energy greater than $0.5 \cdot 10^{10}$ GeV. 
This is due to radiative effects and misidentification 
of the scattered lepton. Approximately 70 events remained in each 
energy interval. The energy depositions from these were averaged 
and compared to the averages from proton showers scaled by 
the ratio of the shower energy to the proton energy. Figure \ref{pnul} 
shows the longitudinal distributions of the hadronic shower energy 
deposited for the different energy intervals (labelled $E_W$) 
compared to the scaled proton distributions. Figure \ref{pnutr} 
shows a sample of the transverse distributions. 

There is a good consistency between the proton and neutrino induced 
showers. The proton showers peak, on average,  20 g cm$^{-2}$ shallower 
in depth with a peak energy $2\%$ larger   
than the neutrino induced showers. This is small compared to the 
overall uncertainty. The slight shift in the longitudinal distribution 
is reflected as a normalisation shift in the radial distributions. 
We conclude therefore that to equate a proton induced shower 
starting at the neutrino interaction point to that 
from a neutrino is a satisfactory approximation. 
  
\section{Parameterisation of showers}

In this section a parameterisation of the energy deposited by the 
showers generated by CORSIKA (averaged over 100 showers depositing 
the same total energy) is described. Other available 
parameterisations will then 
be compared with the showers generated by CORSIKA. 

\begin{figure}
\begin{center}
\includegraphics[height=40pc,width=30pc]{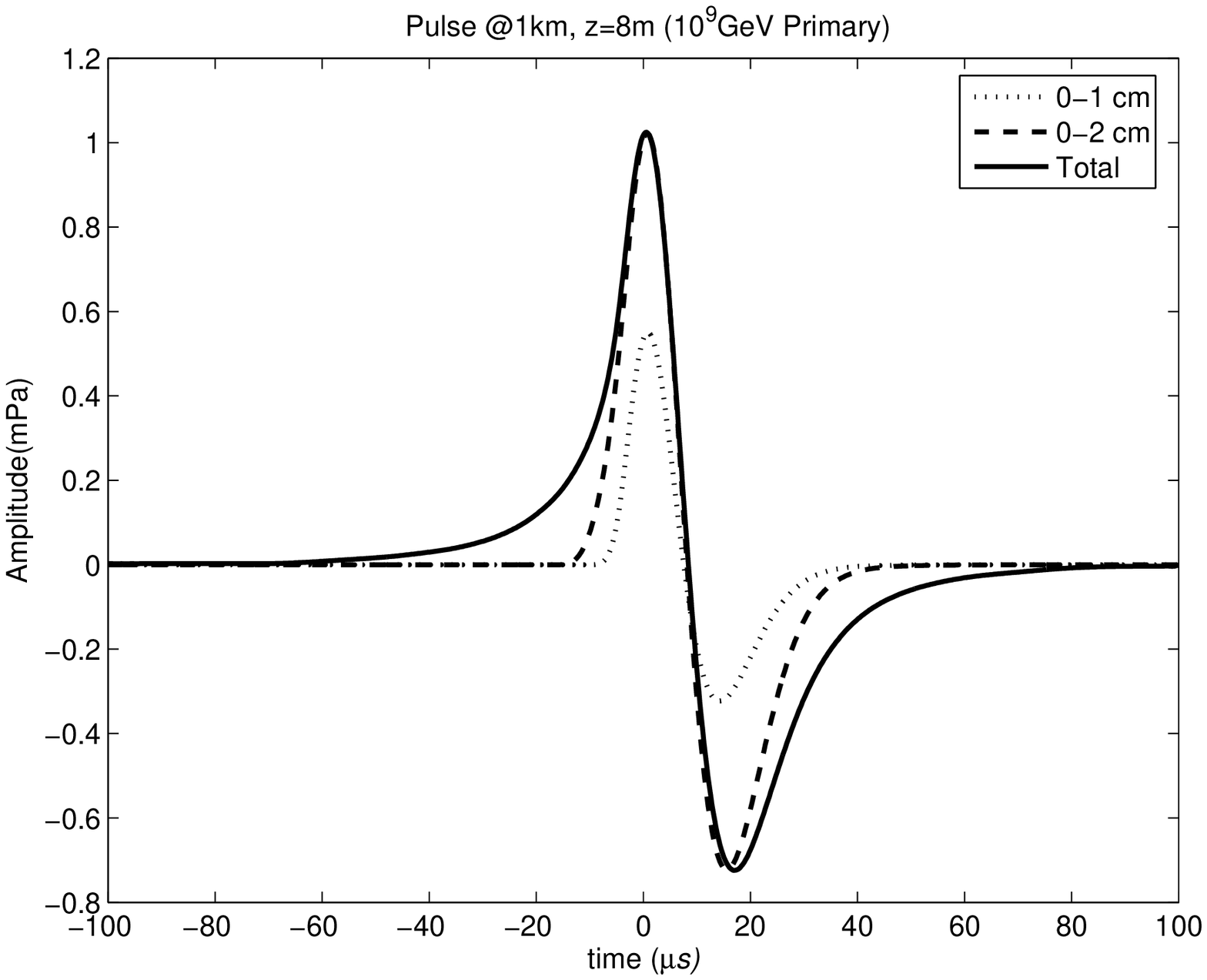}
\vspace*{-5mm}
\caption{\label{build} The acoustic signal at a distance of 1 km from 
the shower axis in the median plane computed from the average 
of 100 CORSIKA showers each depositing a total energy of $10^9$ GeV 
in the water. The dotted, dashed and solid curves shows the signals 
computed from the deposited energies      
within cores of radius 1.025 g cm$^{-2}$, 
2.05 g cm$^{-2}$ and the whole shower (solid curve), respectively.  
It can be seen that most of the amplitude of the signal comes from 
the energy within a core of radius 2.05 g cm$^{-2}$.}
\end{center}
\end{figure}

The acoustic signal generated by a hadron shower depends mainly on 
the energy deposited in the inner core of the shower. This is illustrated 
in figure \ref{build} which shows the contribution to the acoustic 
signal from cores of different radii. This figure shows that it is 
crucial to represent the deposited energy well at radius less than 
2.05 g cm$^{-2}$. The calculation of the acoustic signal 
from the deposited energy is described in section \ref{acoustic}.

\subsection{Parameterisation of the CORSIKA Showers}
\label{ours}

\label{appendix}
The differential energy deposited was parameterised as follows
\begin{equation}
\frac{d^2E}{drdz}=L(z,E_L) \cdot R(r,z,E_L)
\end{equation}
where the function $L(z,E_L)$ represents the longitudinal distribution 
of deposited energy and $R(r,z,E_L)$ the radial distribution. Here $E_L$ 
is $\log_{10} E$ with $E$ the total shower energy. 
   
   The function $L(z,E_L)=dE/dz$ is a modified\footnote{
The modification is to replace the shape parameter $\lambda$ 
in equation 3.5 of reference \cite{Pierre} by the quadratic expression 
in $z$ in equation \ref{Hillas}.} 
version of the Gaisser-Hillas function \cite{Pierre}. This function 
represents the longitudinal distribution of the energy deposited. 
\begin{equation}
L(z,E_L)=P_{1L} \bigg(\frac{z-P_{2L}}{P_{3L}-P_{2L}}\bigg)^
{\frac{(P_{3L}-P_{2L})}{P_{4L}+P_{5L}z+P_{6L}z^2}} \exp\bigg({\frac{P_{3L}-z}{P_{4L}+P_{5L}z+P_{6L}z^2}}\bigg)
\label{Hillas}
\end{equation}
Here the parameters $P_{nL}$ were fitted to quadratic functions 
of $E_L=\log_{10}E$ with values   
\begin{equation}
\frac{P_{1L}}{E}=2.760\cdot 10^{-3} - 1.974 \cdot 10^{-4} E_L + 7.450 \cdot 10^{-6} E_L^2
\label{P1}
\end{equation}
\begin{equation}
P_{2L}=-210.9 - 6.968 \cdot 10^{-3} E_L + 0.1551 E_L^2
\label{P2}
\end{equation}
\begin{equation}
P_{3L}=-41.50 + 113.9 E_L - 4.103 E_L^2
\label{P3}
\end{equation}
\begin{equation}
P_{4L}=8.012 + 11.44 E_L - 0.5434 E_L^2
\label{P4}
\end{equation}
\begin{equation}
P_{5L}=0.7999 \cdot 10^{-5} - 0.004843 E_L + 0.0002552 E_L^2
\label{P5}
\end{equation}
\begin{equation}
P_{6L}=4.563 \cdot 10^{-5} - 3.504 \cdot 10^{-6} E_L +  1,315 \cdot 10^{-7} E_L^2. 
\label{P6}
\end{equation}
The parameter $P_{1L}$ represents the peak energy deposited and $P_{3L}$ the 
depth in the $z$ coordinate at this peak while $P_{2L}$,  
$P_{4L}$, $P_{5L}$ and $P_{6L}$ are related to the shower width and shape 
in $z$. 

The radial distribution was represented by the NKG function \cite{Pierre}  
\begin{equation}
R(r,z,E_L) = \frac{1}{I}\bigg(\big(\frac{r}{P_{1R}}\big)^{(P_{2R}-1)}
\big(1+\frac{r}{P_{1R}}\big)^{(P_{2R}-4.5)}\bigg)  
\end{equation}
where the integral 
\begin{displaymath}
I=\int_0^\infty \bigg(\big(\frac{r}{P_{1R}}\big)^{(P_{2R}-1)}\big(1+\frac{r}{P_{1R}}\big)^{(P_{2R}-4.5)}\bigg)~dr =P_{1R}\frac{\Gamma(4.5-2P_{2R})\Gamma(P_{2R})}{\Gamma(4.5-P_{2R})}. 
\end{displaymath}  
The parameter $P_{1R}$ was found 
to vary strongly with depth while $P_{2R}$ was only a weak function of 
depth. The parameters $P_{nR}$ (with $n=1$,$2$) were each represented by the 
quadratic form 
\begin{equation}
P_{nR}= A + Bz +C z^2 
\label{quadr}
\end{equation}
and the quantities $A,B,C$ parameterised as quadratic 
functions of $E_L$. This gave for $P_{1R}$ 
\begin{equation}
A=0.01287 E_L^2-0.2573 E_L + 0.9636
\end{equation}
\begin{equation}
B=-0.4697 \cdot 10^{-4} E_L^2 + 0.0008072 E_L + 0.0005404
\end{equation}
\begin{equation}
C=0.7344 \cdot 10^{-7} E_L^2-1.375 \cdot 10^{-6} E_L + 4.488 \cdot 10^{-6}
\end{equation}
and for the parameter $P_{2R}$  
\begin{equation}
A= -0.8905 \cdot 10^{-3} E_L^2  + 0.007727 E_L + 1.969 
\end{equation}
\begin{equation}
B= 0.1173 \cdot 10^{-4} E_L^2 - 0.0001782 E_L - 5.093 \cdot 10^{-6}
\end{equation}
\begin{equation}
C=-0.1058 \cdot 10^{-7} E_L^2 + 0.1524 \cdot 10^{-6} E_L - 0.1069 \cdot 10^{-8}.
\end{equation}

The fit was made in a depth range where $dE/dz$ was greater than $10\%$ of 
the peak value defined by equation \ref{P1}. The program MINUIT \cite{minuit} 
was used to minimise the squared fractional deviations 
\begin{equation}
\chi ^2 = \sum_i \bigg( \frac{F_i - D_i}{F_i+D_i} \bigg)^2
\label{chisq}
\end{equation}  
where $F_i$ and $D_i$ refer to the fitted value and the value observed 
in the $i$th bin from  
the CORSIKA showers, respectively. In order to improve the fit at small 
radii the contributions to $\chi^2$ were arbitrarily 
weighted by 10 for $r < 2.05$ g cm$^{-2}$, 4 for $2.05 < r < 3.075$ 
g cm$^{-2}$, unity for $ 3.075 < r < 51.25$ g cm$^{-2}$ and 
0.25 for $r > 51.25 $ g cm$^{-2}$. 
The RMS value of the fractional deviations was $3.4\%$ for radii less 
than 51.25 g cm$^{-2}$ and for energies greater than $10^{6.5}$ GeV. 
The fit becomes poorer at lower energies and greater radii than these. 
Integrating the parameterisation shows that the fraction of the total 
energy computed from the fit within the fit range was $91\%$ 
averaged over the deposited energy range $10^7$ to $10^{12}$ GeV. 
The corresponding fraction directly from the CORSIKA distributions 
was $92.5\%$, averaged over the same energy range. 
When applying this parameterisation at depths with smaller energy 
deposit than $10\%$ of the peak value, the energy was 
assumed to be confined to an annular radius of 1.025 g cm$^{-2}$. 
There was a good agreement (within $5\%$ at the peak) between the 
acoustic signal computed using this parameterisation and that taken 
directly from the CORSIKA showers.

\subsection{The parameterisation used by the SAUND Collaboration}
\label{SAUNDpar}
\begin{figure}
\begin{minipage}{20pc}
\includegraphics[height=50pc,width=21pc]{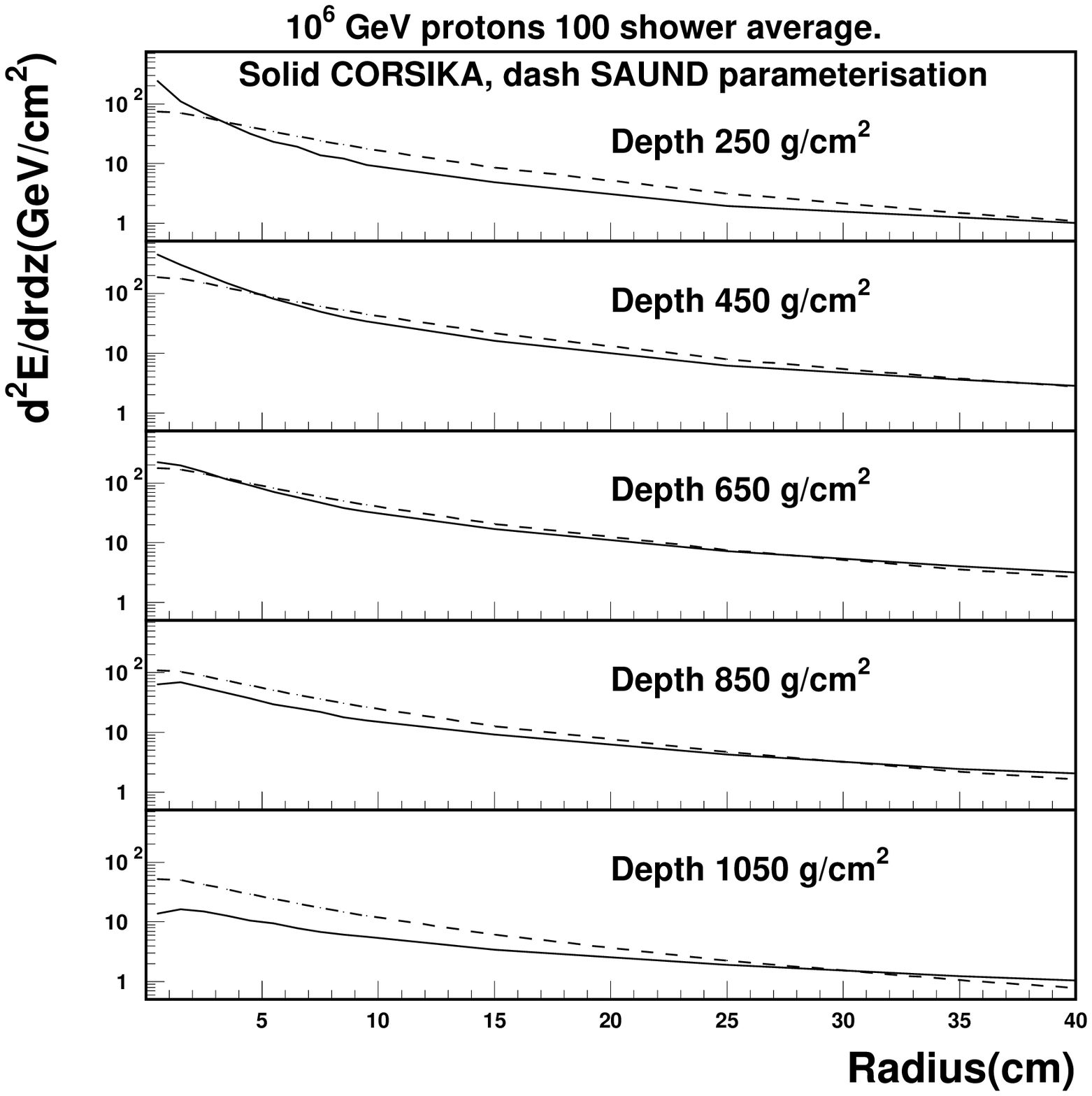}
\end{minipage}\hspace{-1pc}%
\begin{minipage}{20pc}
\includegraphics[height=50pc,width=21pc]{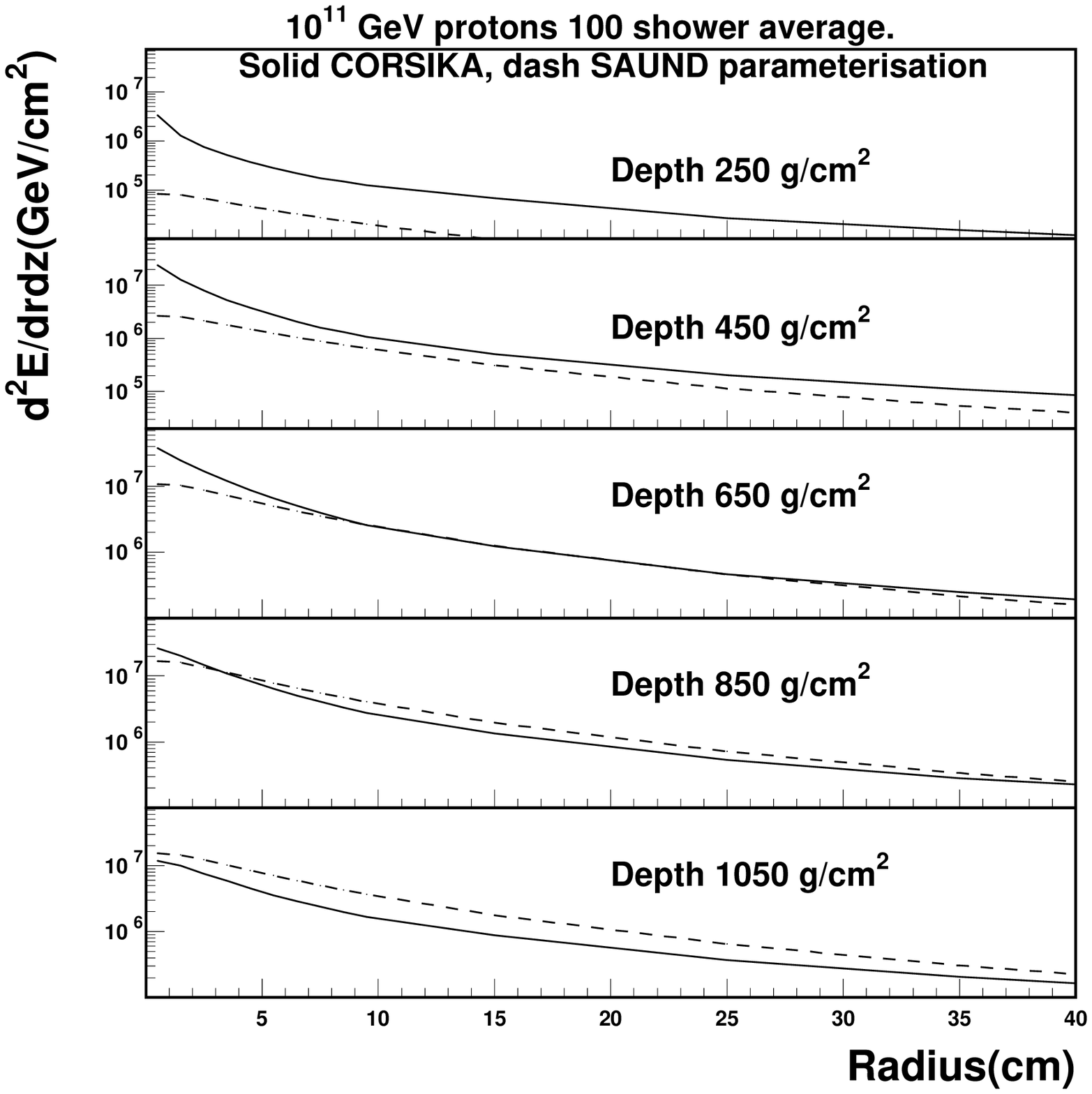}
\end{minipage}\hspace{-1pc}
\vspace*{-5mm}
\caption{\label{nkg}The radial distributions of the deposited energy 
at different depths from CORSIKA compared to the parameterisation used 
by the SAUND collaboration \cite{SAUND} for $10^6$ GeV and $10^{11}$ GeV 
proton induced showers.}
\end{figure}

The SAUND Collaboration \cite{SAUND} uses the following parameterisation
\cite{Justin}, based on the NKG formulae 
(e.g. see reference \cite{Pierre}),  for the energy deposited per unit 
depth, $z$, and per unit annular thickness at radius $r$ from a shower 
of energy $E$
\begin{equation}
\frac{d^2E}{drdz}=E k (\frac{z}{z_{max}})^t \exp{(t-z/\lambda)}~ 2 \pi r \rho(r)
\end{equation}
where $z_{max}=0.9 X_0 \ln(E/E_c)$ is the maximum shower depth,  
$X_0 =36.1$ g cm$^{-2}$ is the radiation length and $E_c = 0.0838$ GeV. 
The constants 
$t=z_{max}/\lambda$ where $\lambda = 130 - 5 \log_{10}(E/10^4 \mathrm{GeV})$ 
g cm$^{-2}$ and $k=t^{t-1}/\exp{(t)}\lambda \Gamma(t)$. 
The radial density is given by  
\begin{equation}
\rho(r)=\frac{1}{r_M^2}a^{s-2} (1+a)^{s-4.5} \frac{\Gamma(4.5-s)}{2\pi \Gamma(s) \Gamma(4.5-2s)}
\label{NKGr}
\end{equation}
where $a=r/r_M$ with $r_M=9.04 $  g cm$^{-2}$, the Moli\`ere radius in water,  
and $s=1.25$.  
Figure \ref{nkg} shows the 
radial distributions from CORSIKA compared with the absolute predictions of 
this parameterisation. 

There is qualitative agreement between the parameterisation and 
the CORSIKA results. The difference in normalisation is explained 
by the somewhat different longitudinal profiles of the CORSIKA showers 
from the SAUND parameterisation. The latter are broader with a lower 
peak energy deposit and a depth of the maximum which is larger than 
the CORSIKA showers. CORSIKA predicts more energy at small 
$r$ than the SAUND parameterisation. Quantitatively, $51\%$ of the shower 
energy is contained within a cylinder of radius 4 cm for the CORSIKA 
showers compared to $35\%$ from the SAUND parameterisation. 
These fractions are approximately independent of energy.  
Hence, in acoustic detectors a harder frequency spectrum for 
the acoustic signals is predicted by CORSIKA than by the SAUND 
parameterisation. Note that in the fit described in Section \ref{appendix}  
the values of the parameter $P_{1R}$ (equivalent to $R_M$ in equation
\ref{NKGr}) were strongly depth dependent 
and much lower than the Moli\`ere radius in water, assumed by the SAUND 
collaboration. In addition, the value of $P_{2R}$ 
(equivalent to $s$ in equation \ref{NKGr}) while relatively 
constant tended to be at a higher value ($\sim 1.9$) than that 
assumed by SAUND.      

\subsection{The parameterisation used by Niess and Bertin}
Hadron showers, generated by Geant4 (version 4.06 p03), were studied up 
to energies of $10^5$ GeV and electromagnetic showers to higher energies 
by Niess and Bertin \cite{BN,Valentin}. The hadronic showers were 
parameterised as follows.   
\begin{equation}
\frac{d^2E}{drdz}=r f(z) g(r,z)
\end{equation}
with 
\begin{equation}
f(z) = \frac{E}{X_0} b \frac{(b z^\prime)^{a-1} \exp{-bz^\prime}}{\Gamma(a)}
\end{equation}
where $E$ is the energy of the hadron shower, $X_0$ is the radiation 
length in water, $z^\prime = z/X_0$, $b=0.56$ as 
determined from the fit and $a$ is chosen to satisfy 
$z_{max}^\prime = (a-1)/b$. 
Here $z_{max}^\prime$ is the depth in radiation lengths at which 
the shower maximum occurs. This is parameterised as 
\begin{equation}
z_{max}^\prime = 0.65 \log (\frac{E}{E_c}) + 3.93  
\end{equation}  
with $E_c=0.05427$ GeV. The radial distribution function is parameterised as 
\begin{equation}
g(r,z)=g_0 \bigg(\frac{r_i}{r}\bigg)^n
\end{equation} 
where $r_i=3.5$ cm, $n=n_1=1.66-0.29(z/z_{max})$ for $r < r_i$ and 
$n=n_2=2.7$ for $r>r_i$. The constant $g_0$ is chosen to be 
$(2-n_1)(n_2-2)/((n_2-n_1)r_i^2)$ so that the  integral of the radial 
distribution is unity. 

\begin{figure}
\begin{minipage}{20pc}
\includegraphics[height=50pc,width=21pc]{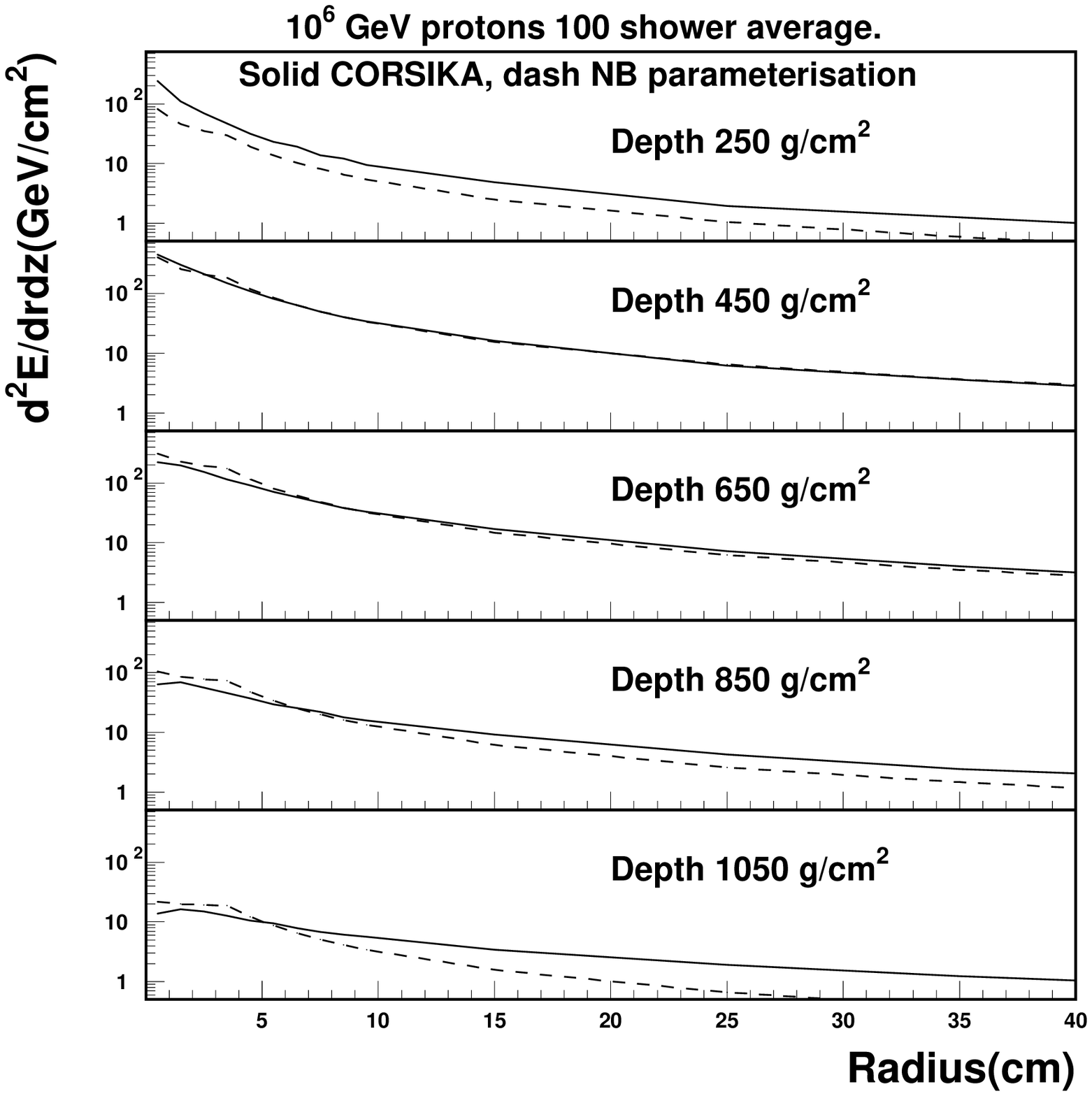}
\end{minipage}\hspace{-1pc}%
\begin{minipage}{20pc}
\includegraphics[height=50pc,width=21pc]{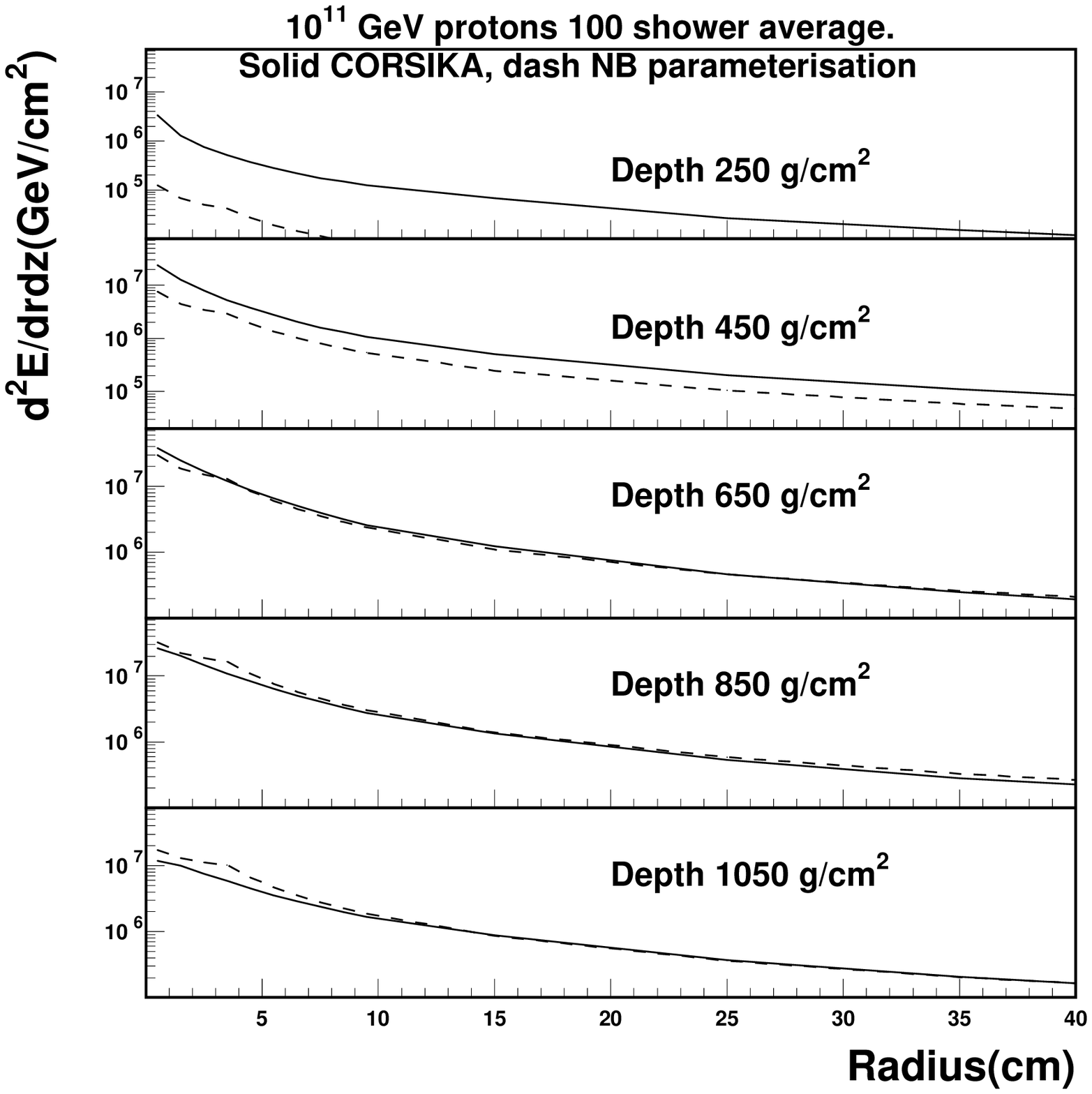}
\end{minipage}\hspace{-1pc}
\vspace*{-5mm}
\caption{\label{BNr}The radial distributions of the deposited energy 
at different depths from CORSIKA compared to the parameterisation used 
by the Niess and Bertin \cite{Valentin,BN} (labelled NB parameterisation) 
for $10^6$ GeV and $10^{11}$ GeV proton induced showers.}
\end{figure}

\begin{figure}[h]
\includegraphics[width=36pc]{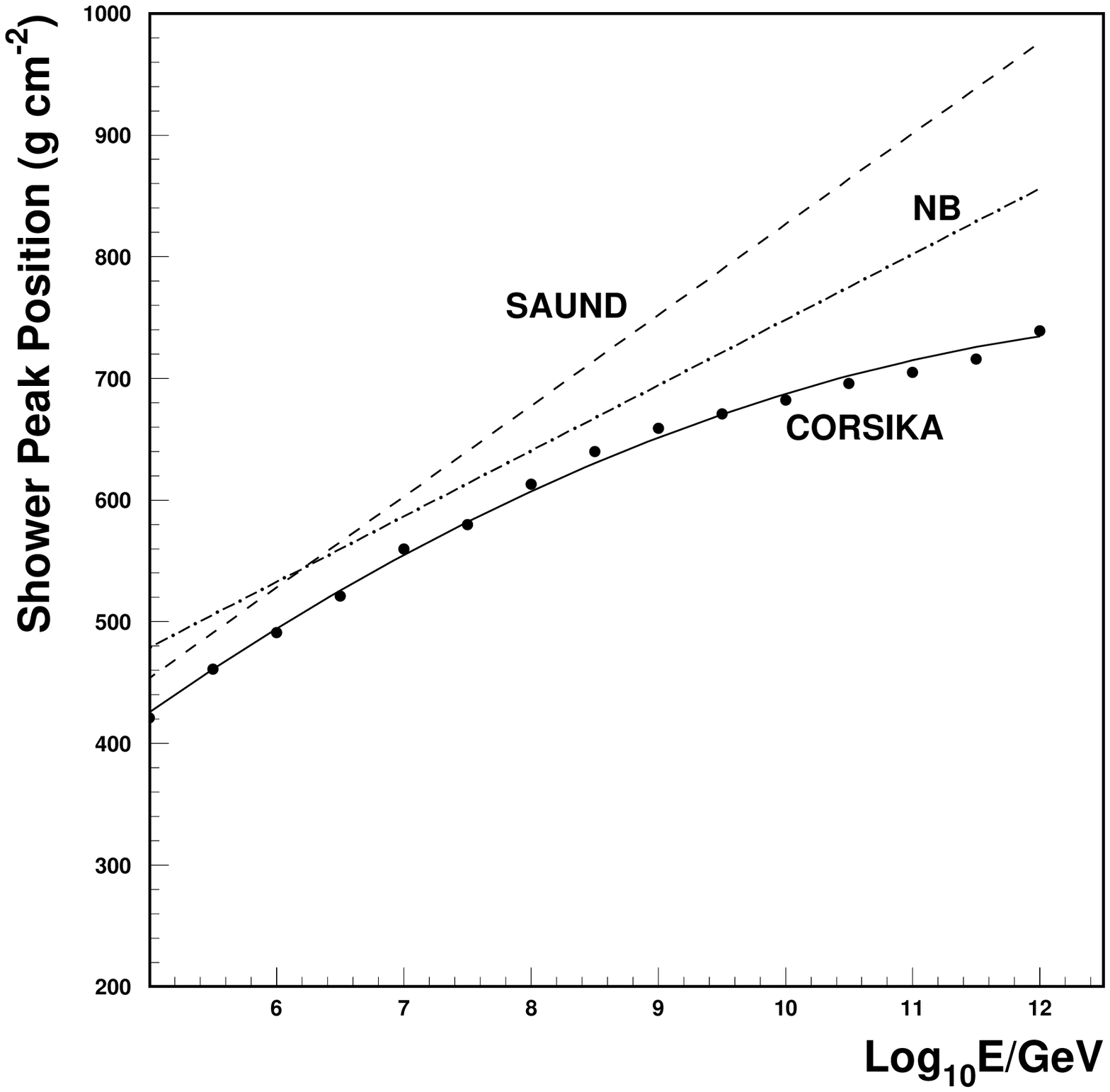}\hspace{2pc}%
\caption{\label{longP3}The depth of the shower peak as a function of 
$\log_{10} E$ from CORSIKA (black points) for the showers starting at $z=0$. 
The solid curve shows the parameterisation according 
to equation \ref{P3}. The dashed (dash dotted) lines show the values 
assumed by the SAUND (Niess-Bertin) Collaborations.}
\end{figure}

Figure \ref{BNr} shows the radial distributions from this parameterisation 
compared with the predictions of CORSIKA. There is quite good agreement 
between the two. There is a difference in the normalisation with depth 
since Geant4, on which this parameterisation is based, produces showers 
which tend to develop more slowly with depth than those from CORSIKA 
(see Figure \ref{long}). Furthermore, both this and the SAUND 
parameterisation (Section \ref{SAUNDpar}) assume a linear variation of 
the shower peak depth with $\log E$ whereas CORSIKA gives a clear parabolic 
shape (see equation \ref{P3}). This is illustrated in Figure \ref{longP3}. 
The Niess-Bertin parameterisation predicts that $56\%$ of the shower 
energy is contained within a cylinder of radius 4 cm in reasonable 
agreement with the value of $51\%$ 
from CORSIKA (these values are almost independent of energy). 

\section{The acoustic signals from the showers.}
\label{acoustic} 

The pressure, $P$, from a hadron shower depositing total energy $E$ at 
time $t$ resulting from the deposition of relative energy density 
$\epsilon = (1/E)(1/2\pi r)d^2E/drdz$ at 
a point distant $d$ from the volume, $dV$, follows the form \cite{Learned}
\begin{equation}
P(d,t) = \frac{E\beta}{4\pi C_p} \int \frac{\epsilon}{d}\frac{d}{dt}\bigg(\delta(t-d/c)\bigg) dV 
\label{pdel}
\end{equation}
where the integral is over the total volume of the shower. Here 
$\beta=2.0\cdot10^{-4}$ is the thermal expansion 
coefficient of the medium at $14^\circ$C, $C_p=3.8\cdot 10^{3}$ 
J kg$^{-1}$ K$^{-1}$ is the 
specific heat capacity and $c=1500$ ms$^{-1}$ is the velocity of 
sound in the sea water.  

Acoustic signals seen by an observer at distance $r$ from the shower centre 
are computed from equation (\ref{pdel}) as follows.  
Points are produced randomly throughout the volume of the shower with 
density proportional to the deposited energy density and the time of 
flight from every produced point to the observer calculated. The flight 
times to the observer are histogrammed over $2^n$ bins (in this case 
$n=10$ is chosen) centred on the mean flight time and with a suitable 
bin width, $\tau$ (chosen here to be 1$\mu$s). The counts in each bin 
of the histogram are divided by $\tau$ yielding the function $E_{xyz}(t)$. 
The Fourier transform of the pressure wave is then 
\begin{equation} 
P(\omega) = \frac{1}{r} \int_{-\infty}^\infty \frac{E \beta}{4 \pi C_p} \frac{d}{dt}E_{xyz}(t)e^{-i \omega t} dt = \frac{1}{r} \frac{E \beta}{4 \pi C_p} i \omega \int_{-\infty}^\infty E_{xyz}(t) e^{-i \omega t} dt = \frac{1}{r} \frac{E \beta}{4 \pi C_p} i \omega E_{xyz}(\omega)  
\end{equation} 
using the standard Fourier transform theorem, that taking the derivative 
in the time domain is the same as multiplying by $i\omega$ in the frequency 
domain. The Fourier transform $E_{xyz}(\omega)$ at 
angular frequency $\omega$ is evaluated numerically by a fast Fourier 
Transform (FFT) from the histogram 
$E_{xyz}(t)$. A correction is applied for attenuation in the water by   
a factor $A(\omega)=e^{-\alpha(\omega) r}$ where $\alpha(\omega)$ is 
the frequency dependent attenuation coefficient.  
The pressure as a function of time is then evaluated numerically 
by an inverse FFT using frequency steps from zero to the 
sampling frequency (the inverse of the bin width $\tau$ i.e. 1 MHz 
in this case). This gives  
\begin{equation}
P(t)= \frac{1}{1024}\sum_{n=-512}^{n=511} P(\omega_n) A( \omega_n ) e^{i n \Omega}
\end{equation} 
where $\Omega=2 \pi/1024$ radians and $\omega_n/2 \pi = n \Omega /2\pi$ MHz 
is the $n$th frequency.  
The attenuation coefficient $\alpha ( \omega )$ is computed either  
according to the formulae in \cite{AM} or using the complex attenuation  
given in \cite{BN,Valentin}.   
This method of calculation was computationally much faster than the  
evaluation of the space integral given in equation 18 of reference 
\cite{Learned} and gave identical results.  

\begin{figure}
\begin{minipage}{20pc}
\hspace{-3pc}\includegraphics[height=21pc,width=21pc]{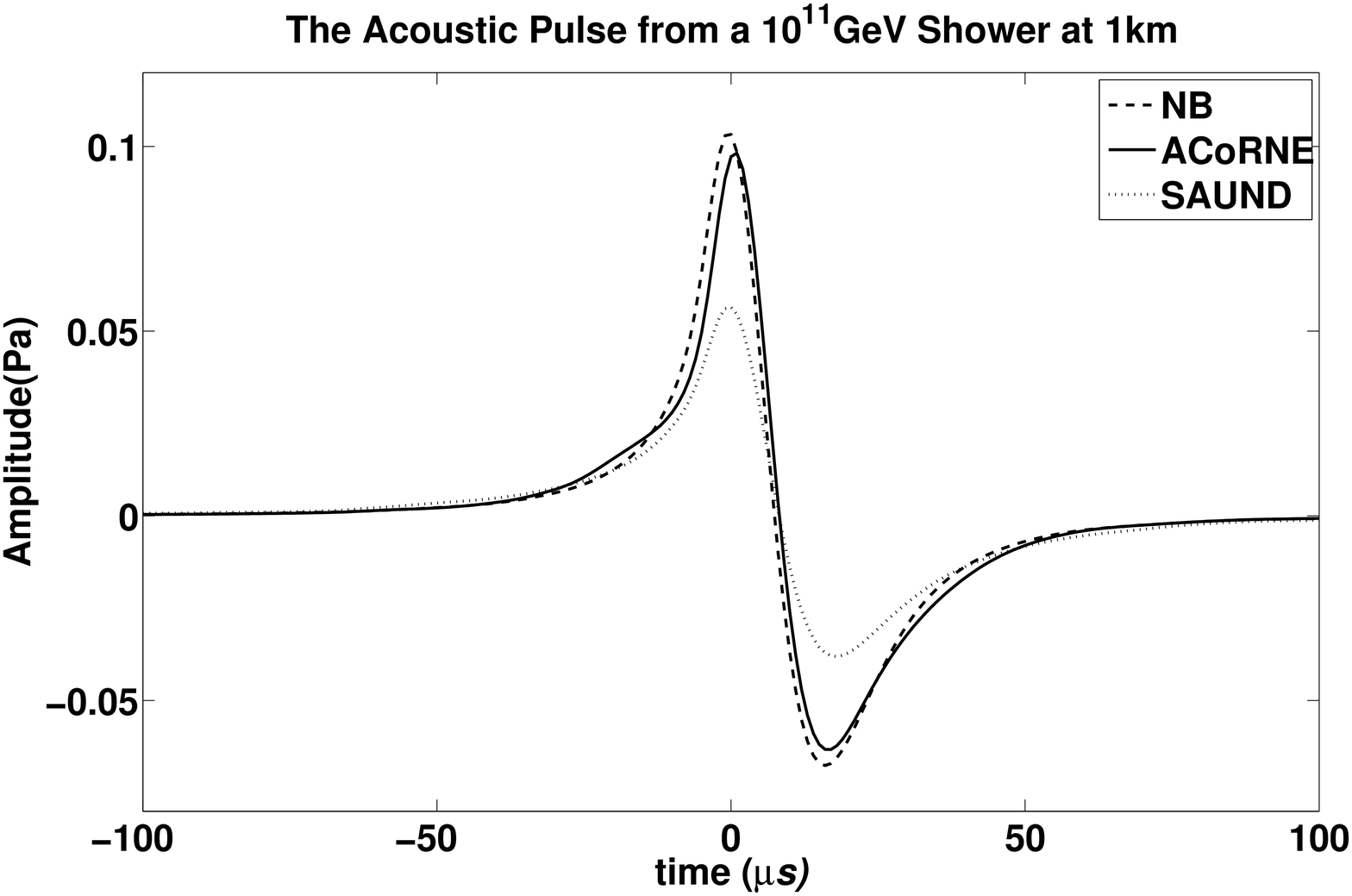}
\end{minipage}\hspace{-1pc}%
\begin{minipage}{20pc}
\includegraphics[height=21pc,width=21pc]{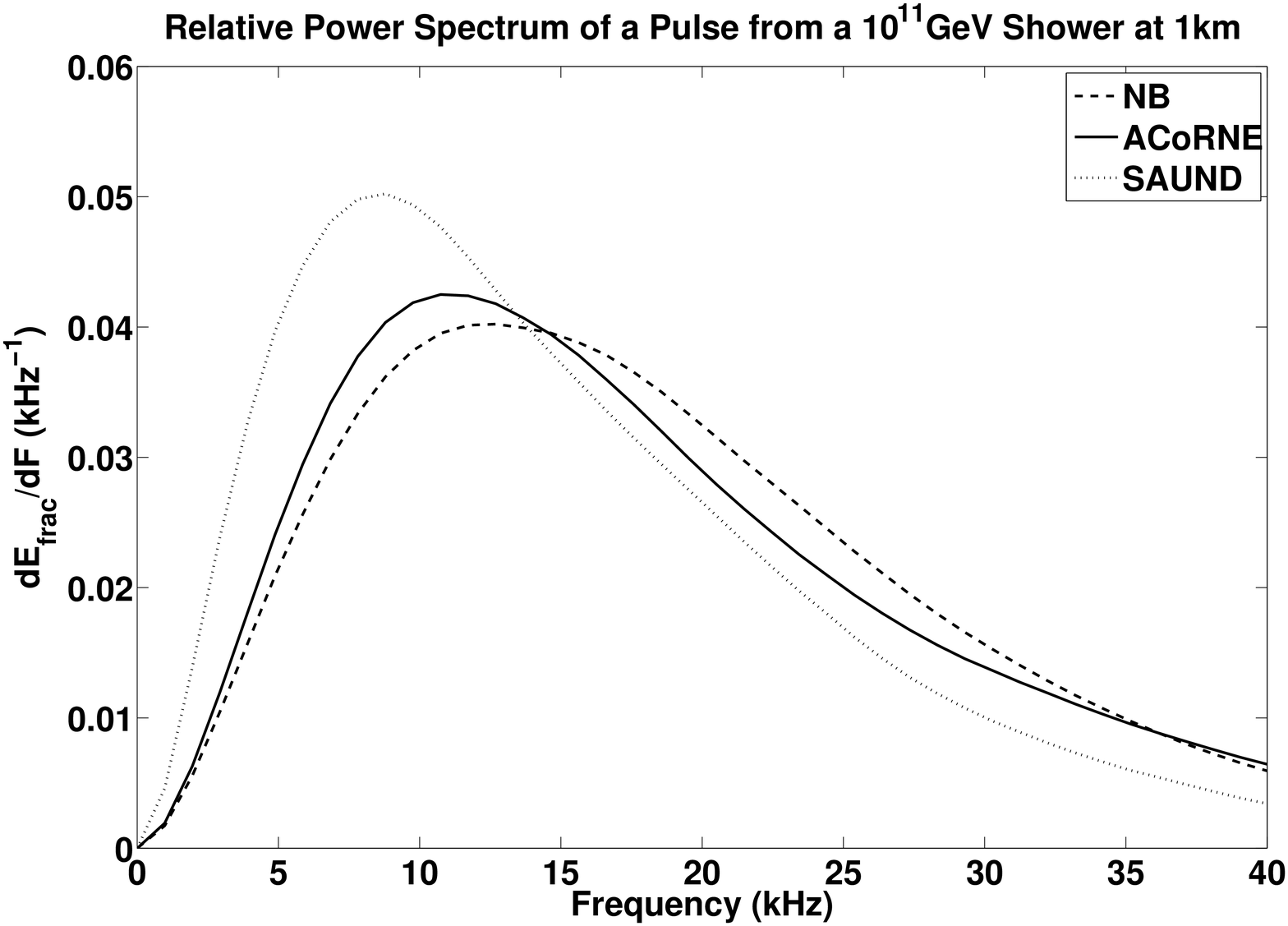}
\end{minipage}\hspace{-1pc}
\vspace*{-5mm}
\caption{\label{pulses}The left hand plot shows acoustic pulses generated 
from the parameterisation described in section \ref{ours} labelled ACoRNE,  
the parameterisation from reference \cite{BN,Valentin} labelled NB and 
that from reference \cite{SAUND,Justin} labelled SAUND. 
These pulses were evaluated for a hadron shower from a neutrino 
interaction depositing hadronic energy of $10^{11}$ GeV 1 km distant 
from an acoustic detector in a plane perpendicular to the shower 
axis at the shower maximum (the median plane). The right hand 
plot shows the frequency decomposition of the pulses in the left 
hand plot.}
\end{figure}

Acoustic pulses, computed with the complex attenuation described 
in \cite{BN,Valentin}, 
using the parameterisations of the shower profile given 
above are shown in Figure \ref{pulses}. It can be seen that the 
parameterisation developed here gives similar results to that 
described in \cite{BN,Valentin} despite the fact that the latter was 
an extrapolation from low energy simulations. The parameterisation 
used by SAUND \cite{SAUND,Justin} gives smaller signals 
concentrated at somewhat lower frequencies.

\begin{figure}
\begin{minipage}{20pc}
\hspace{-3pc}\includegraphics[height=21pc,width=21pc]{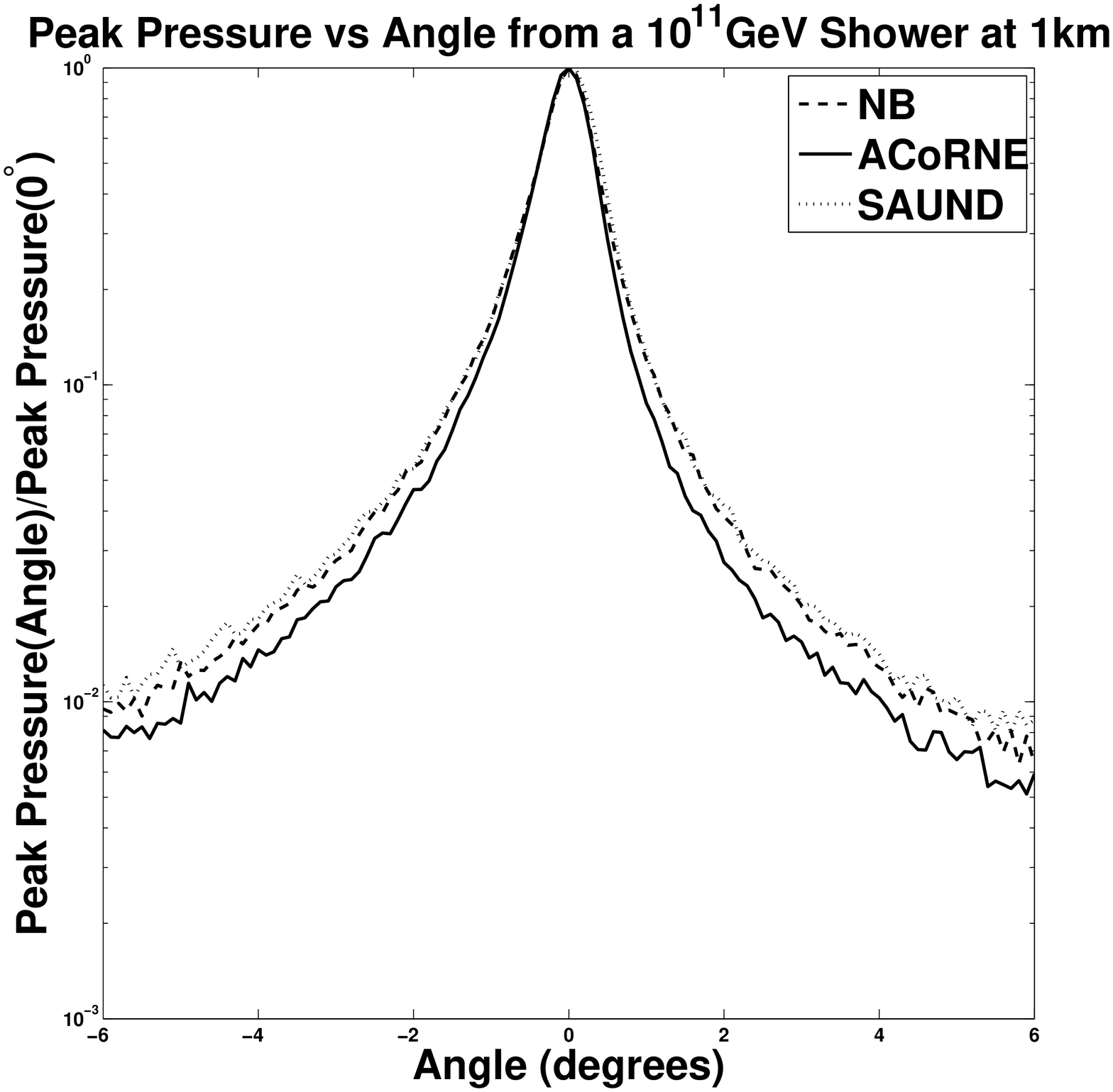}
\end{minipage}\hspace{-1pc}%
\begin{minipage}{20pc}
\includegraphics[height=21pc,width=21pc]{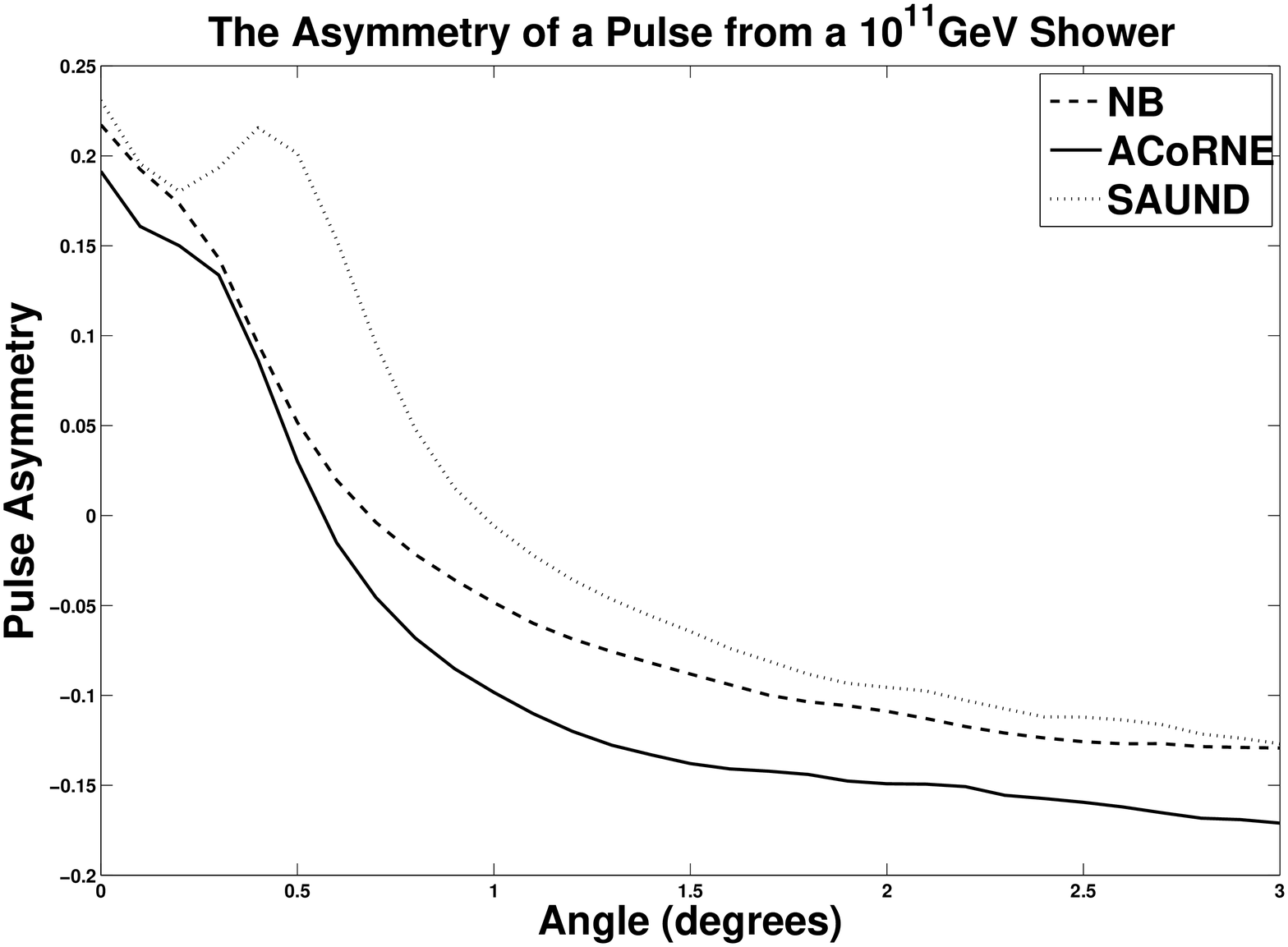}
\end{minipage}\hspace{-1pc}
\vspace*{-5mm}
\caption{\label{angles}The left hand plot shows the variation of the 
peak pressure in the pulse with angle from the median plane   
at 1 km from the shower. The right hand plot shows the 
variation of the pulse asymmetry with this angle at the same distance.
The curves were computed from the parameterisations labelled.}
\end{figure}

\begin{figure}
\begin{minipage}{20pc}
\hspace{-3pc}\includegraphics[height=21pc,width=21pc]{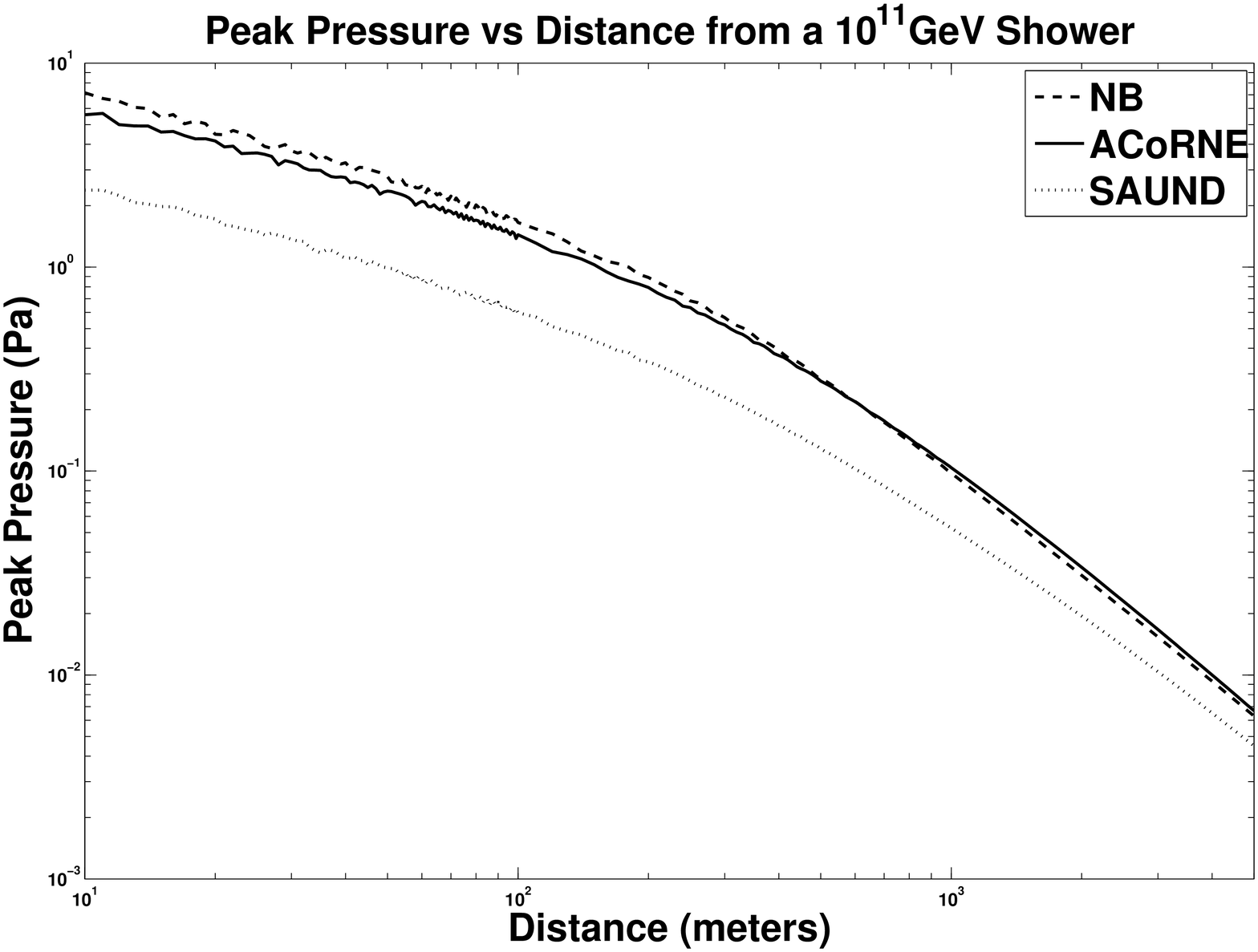}
\end{minipage}\hspace{-1pc}%
\begin{minipage}{20pc}
\includegraphics[height=21pc,width=21pc]{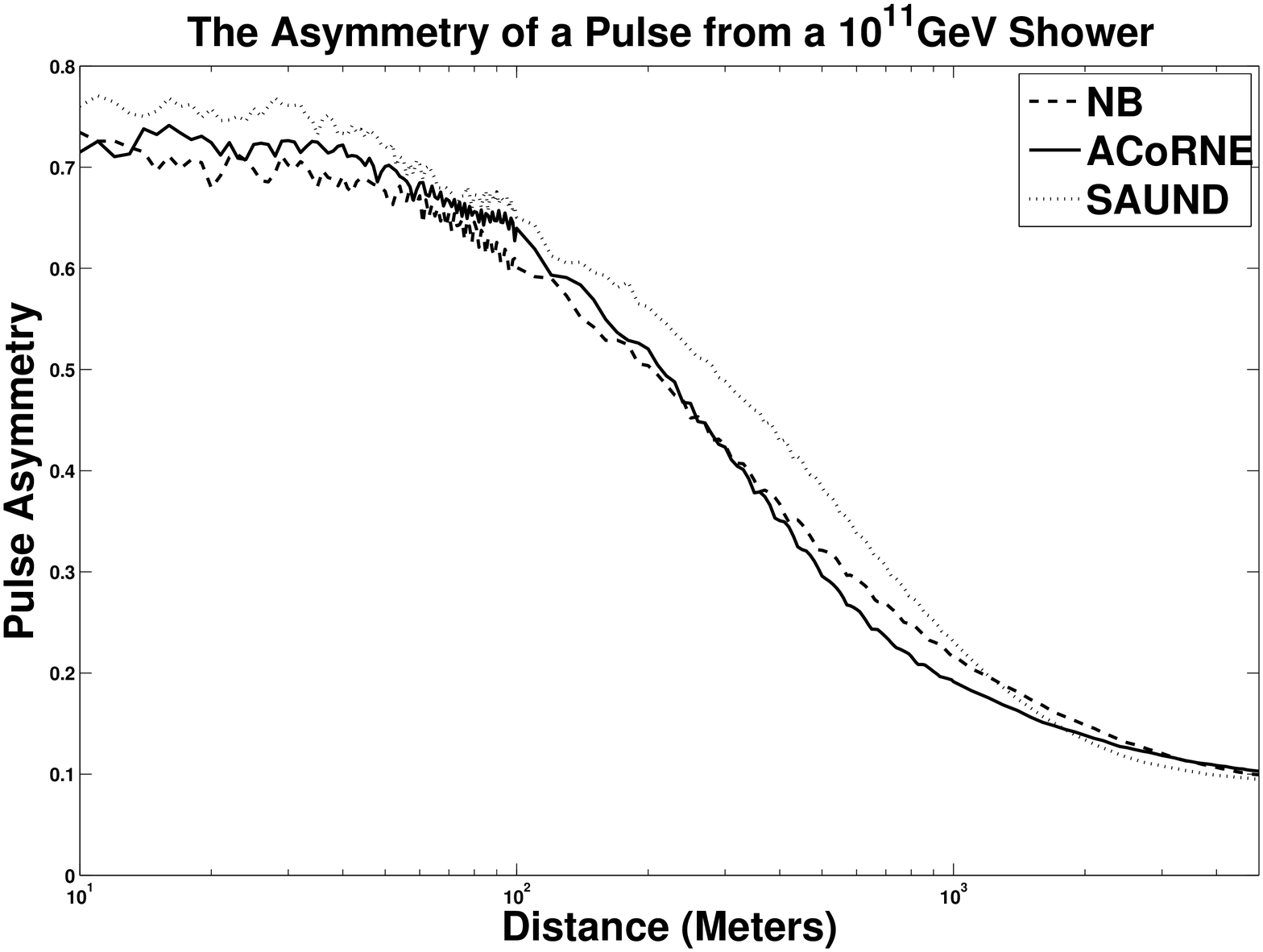}
\end{minipage}\hspace{-1pc}
\vspace*{-5mm}
\caption{\label{dists}The left hand plot shows the decrease of the pulse 
peak pressure and the right hand plot the pulse asymmetry, both in the median 
plane, as a function of distance from the shower computed from the 
parameterisations.}
\end{figure}

\begin{figure}
\begin{minipage}{20pc}
\hspace{-3pc}\includegraphics[height=21pc,width=21pc]{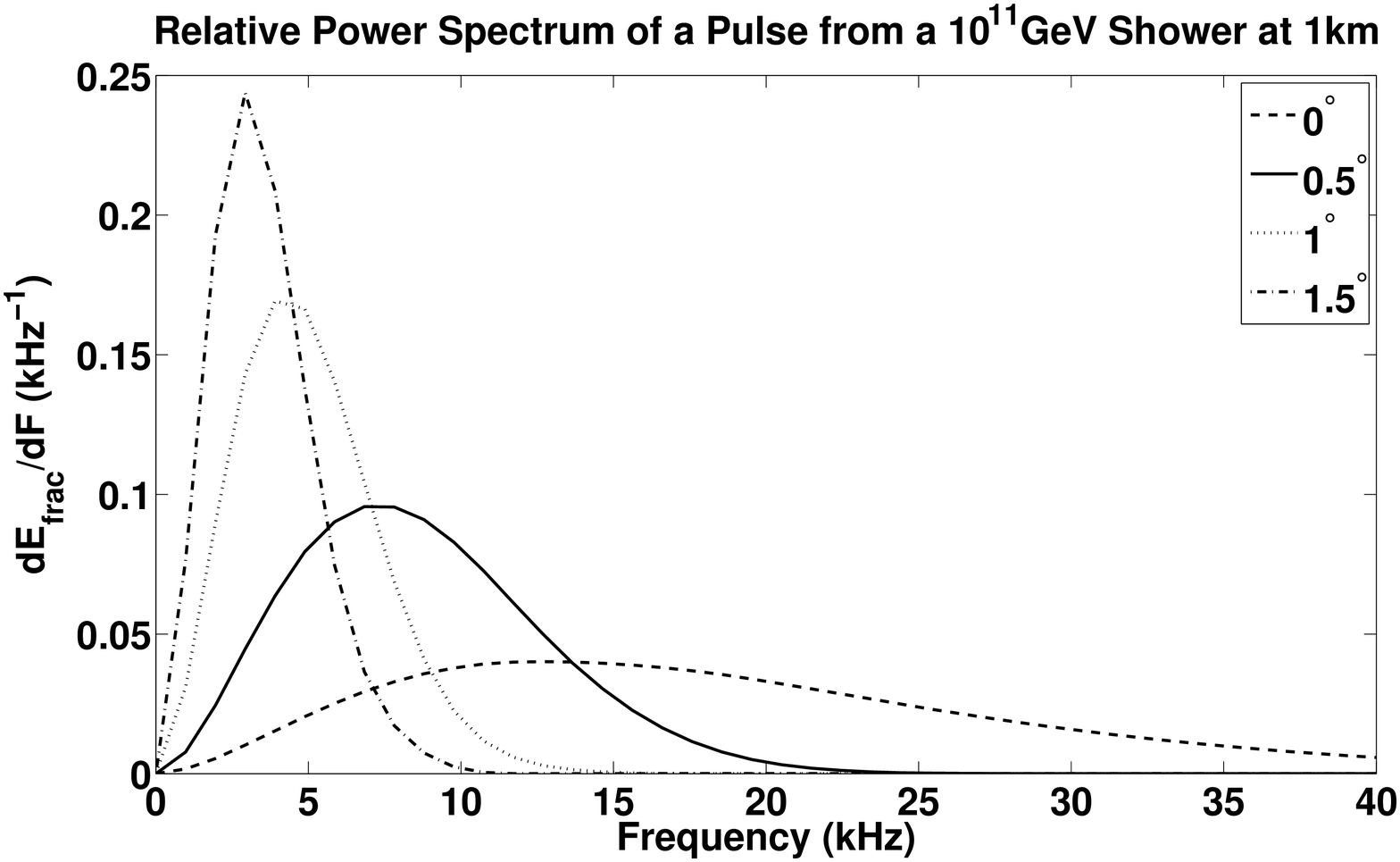}
\end{minipage}\hspace{-1pc}%
\begin{minipage}{20pc}
\includegraphics[height=21pc,width=21pc]{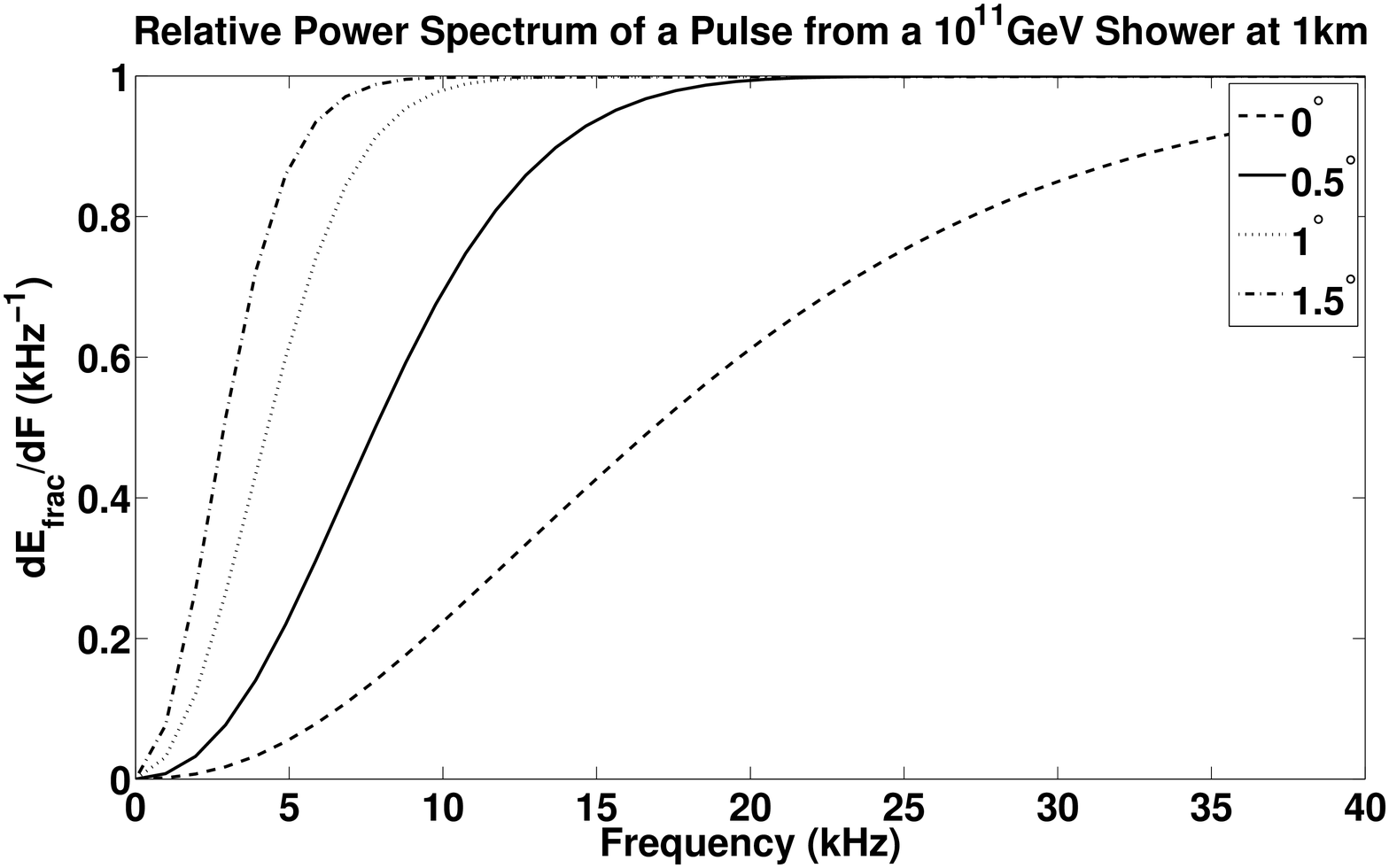}
\end{minipage}\hspace{-1pc}
\vspace*{-5mm}
\caption{\label{eangle}The left hand plot shows the frequency decomposition 
of the acoustic signal, computed from the parameterisation of the CORSIKA 
showers, at different angles to the median plane at a distance 
of 1km from the shower and the right hand plot shows the cumulative 
frequency spectrum i.e. the integral of the left hand plot. }
\end{figure}

\begin{figure}
\begin{minipage}{20pc}
\hspace{-3pc}\includegraphics[height=21pc,width=21pc]{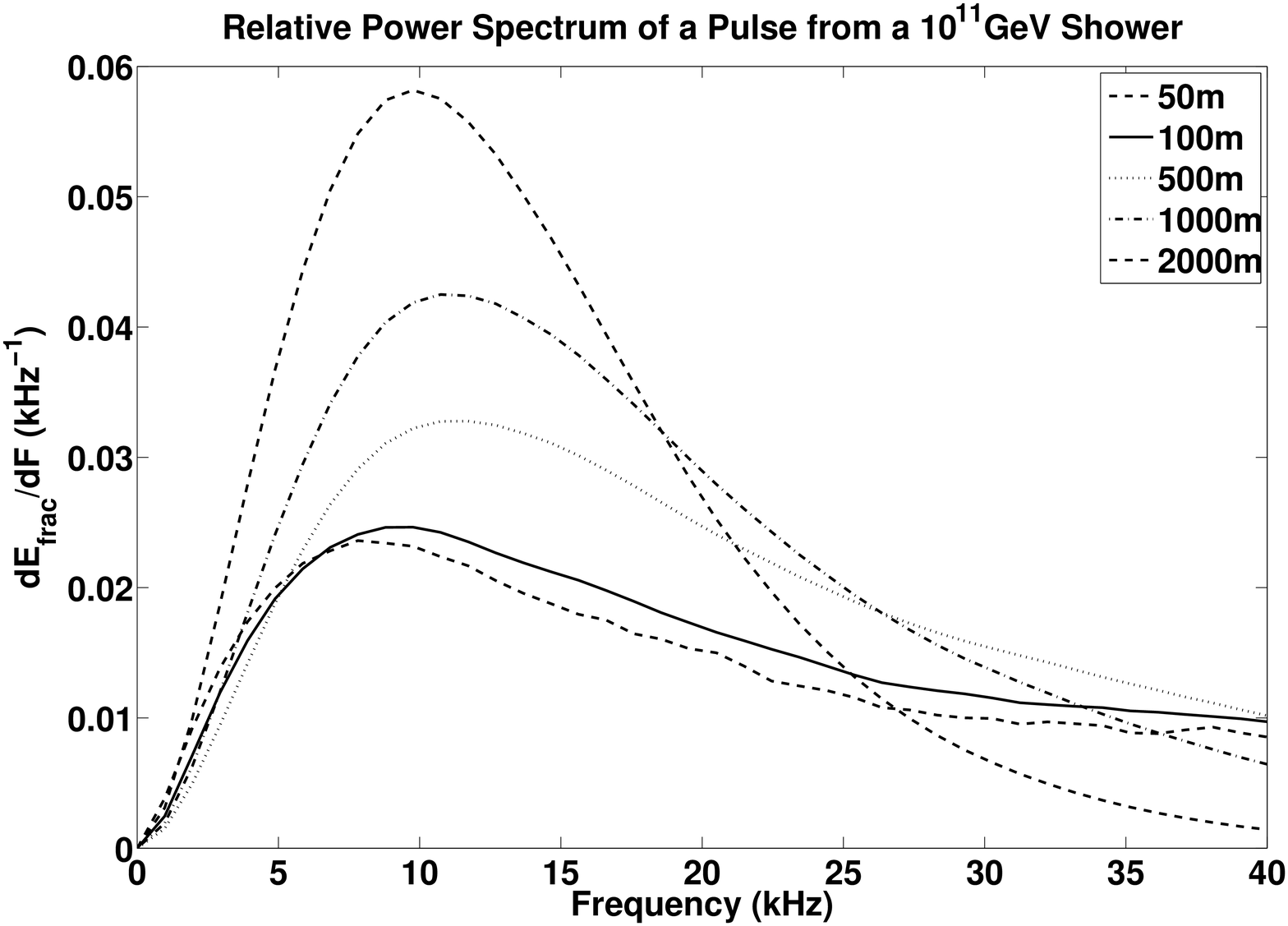}
\end{minipage}\hspace{-1pc}%
\begin{minipage}{20pc}
\includegraphics[height=21pc,width=21pc]{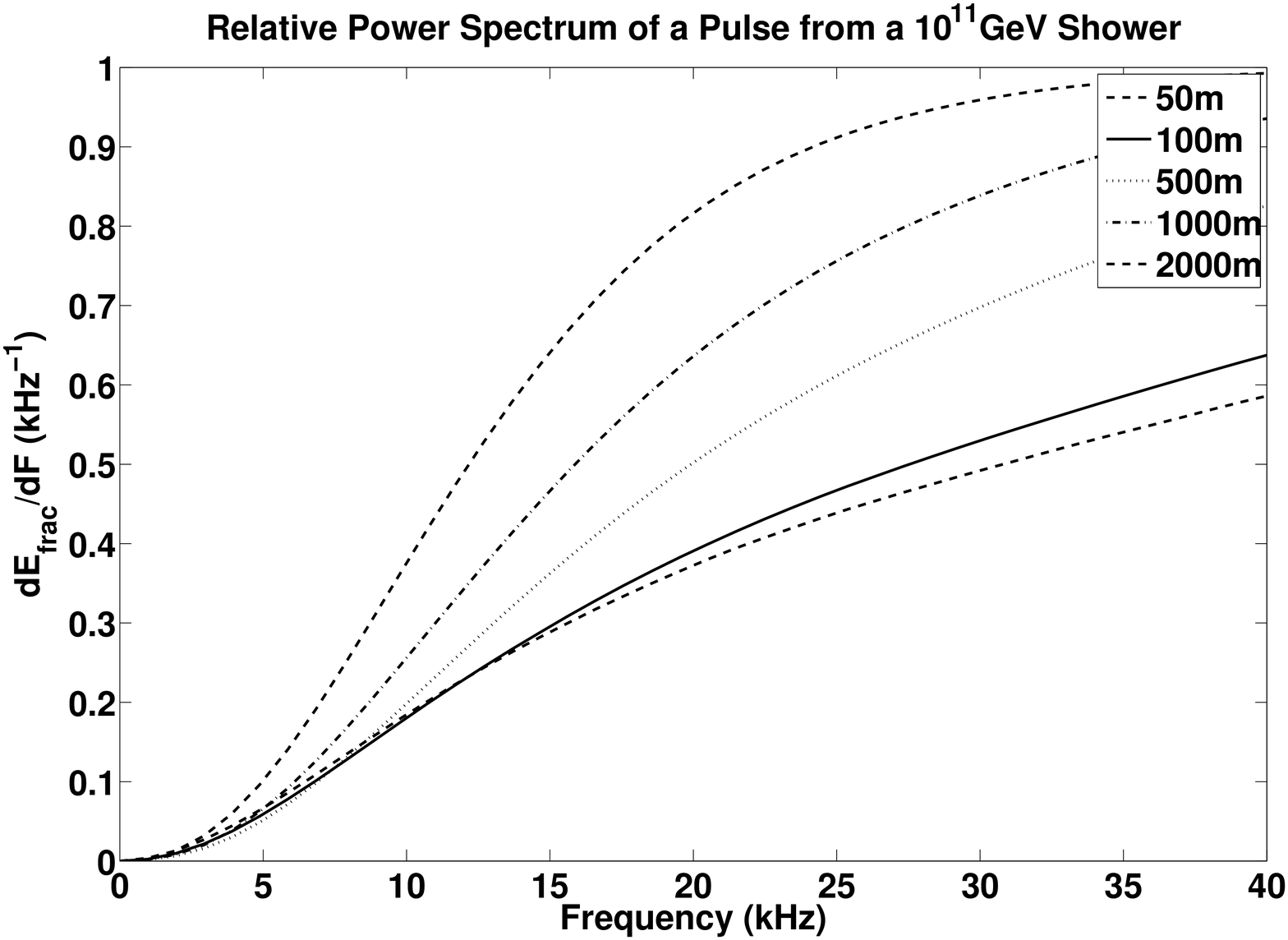}
\end{minipage}\hspace{-1pc}
\vspace*{-5mm}
\caption{\label{edist}The left hand plot shows the frequency decomposition 
of the acoustic signal, computed from the parameterisation of the CORSIKA 
showers, at different distances from the hydrophone in the 
median plane and the right hand plot shows the cumulative 
frequency spectrum i.e. the integral of the left hand plot. }
\end{figure}

Further properties of the acoustic signals are shown in Figures 
\ref{angles} to \ref{edist}. The pulses tend to be somewhat asymmetric 
with the asymmetry defined by $|P_{max}|-|P_{min}|/|P_{max}|+|P_{min}|$. 
The complex nature of the attenuation enhances this asymmetry. 
This is most evident in the far field conditions e.g. at 5km 
where non complex attenuation would yield a totally symmetric pulse.  
Figure \ref{angles} shows the angular dependence of the peak pressure.  
Here the angle is that subtended by the acoustic detector relative 
to the plane, termed the median plane,  through the shower maximum 
at right angles to the axis of the shower. The parameterisation 
derived here gives a somewhat narrower angular spread than the others. 
This could be due to the slightly longer showers predicted by CORSIKA 
than the others. Figure \ref{angles} also 
shows the asymmetry of the pulse as a function of this angle. The 
pulse initially becomes more symmetric moving out of the median plane 
and then the asymmetry becomes negative at larger angles.  
Figure \ref{dists} shows the decrease of the pulsed peak pressure 
with distance from the shower in the median plane and the asymmetry 
with distance in this plane. Figures \ref{eangle} and \ref{edist} 
show the frequency composition of the pulses at different angles 
to the median plane at 1 km from the shower 
and at different distance in the median plane, respectively.

\section{Conclusions}
  
The simulation program for high energy cosmic ray air showers, CORSIKA, 
has been modified to work in a water or ice medium. This allows both 
hadron and neutrino showers to be generated in the medium over a wide 
range of energy ($10^5$ to $10^{12}$ GeV). The properties of 
hadronic showers in water simulated by CORSIKA agree with those from other 
simulations to within $10-20\%$. A similar uncertainty has been noted 
previously from the variations in CORSIKA showers in air generated by 
different models of the hadron interactions. However, none of the other 
available simulations for water cover the range of energies accessible 
to CORSIKA. The hadronic showers produced by neutrino interactions are 
shown to have similar profiles to proton showers which deposit the same 
amount of energy to that from the neutrino and which start at the 
interaction point of the neutrino. The properties of the neutrino 
interactions are described.   
A parameterisation of the shower profiles generated by CORSIKA is given.    
There is reasonable agreement with the parameterisation based on the 
Geant4 simulations at low energy ($<10^5$ GeV) developed by Niess and 
Bertin. However, the agreement with the parameterisation used by the 
SAUND Collaboration, which is based on the NKG formalism, is less good.   
The position of the shower maximum, determined from the CORSIKA program, 
is found to vary quadratically with $\log E$ rather than linearly  
as assumed in the latter two parameterisations. 

The acoustic signals generated by neutrino interactions using 
CORSIKA and by the two other parameterisations are described 
and their properties are studied. The acoustic signal is found to 
be very sensitive to the energy deposited close to the shower axis. 

\subsection{Acknowledgments}
We wish to thank Ralph Engel, Dieter Heck, Johannes Knapp and Tanguy 
Pierog for their assistance in modifying the CORSIKA program. We also 
thank Valentin Niess and Justin Vandenbroucke for valuable discussions.  

\vspace*{-1mm}

\vspace*{-1mm}

\vspace*{-1mm}

\vspace*{-1mm}

\vspace*{-1mm}

\vspace*{-1mm}

\vspace*{-1mm}

\vspace*{-1mm}

\vspace*{-1mm}

\vspace*{-1mm}



\begin{thebibliography}{99}

\bibitem{ARENAs} Proceedings of the Workshop on Acoustic and 
Radio EeV Neutrino Detection Activities (ARENA), DESY, Zeuthen (May 2005),  
Editors R. Nahnhauer and S. B\"oser

\bibitem{GZK}K. Griesen, Phys. Rev. Lett.16 (1966) 748, \\
G.T. Zaptsepin, V.A. Kuzmin, JETP Lett. 4 (1966) 78. 

\bibitem{WBlim} E. Waxman and J. Bahcall, Phys. Rev. D59 (1999) 023002,  
(hep-ph/9807282). 

\bibitem{GZKnu} R.D. Engel, D. Seckel and T.Stanev, 
Phys. Rev. D64 (2001) 093010 (astro-ph/0101216).  

\bibitem{AMANDA} See for example M. Ackermann et al., (astro-ph/0412347) 
Phys. Rev. D71 (2005) 077102. 

\bibitem{IceCube} See for example Nucl. Instrum. Meth. A567 (2006) 438 

\bibitem{Antares} J.A. Aguilar et al., astro-ph/0606229.

\bibitem{NESTOR} See for example G. Aggouras et al., 
Nucl. Instrum. and Meth. A552 (2005) 420. 

\bibitem{Auger} See for example Nucl. Phys. Proc. Suppl. 143 (2005) 373.  

\bibitem{CORSIKA} ``CORSIKA: A Monte Carlo Code to Simulate Extensive Air 
Showers'', D. Heck et al., Karlsruhe Report FZKA 6019.  
(http://www-ik.fzk.de/corsika). 


\bibitem{Asky} G. Askar'yan, Soviet Physics JETP 14 (1962) 441 and 
21 (1965) 658. 

\bibitem{JZ} J. Alvarez-Mu\~niz, E. Marqu\'es, 
R.A. V\'azquez and E. Zas Phys. Rev. D68 (2003) 043001 
(astro-ph/0206043). 

\bibitem{Learned} J.G. Learned Phys. Rev. D19 (1979) 3293.

\bibitem{Zasetal} J. Alvarez-Mu\~niz and E.Zas, Phys. Lett. B434 (1998) 396 
(astro-ph/9806098).

\bibitem{BN} V. Niess and V. Bertin astro-ph/0511617 and 
V. Niess, PhD Thesis, CPPM, Marseille.

\bibitem{Valentin} V. Niess, PhD Thesis, CPPM, Marseille, see equations 
1-55 and 1-56.

\bibitem{Geant4} Geant4, J. Allison et al., 
{\it Nucl. Inst. and Meths. in Phys. Research} {\bf A506} (2003) 250 and 
{\it IEEE Transactions on Nucl. Science} {\bf 53} (2006) 270.

\bibitem{EGS} ``The EGS4 Code System'' W.R. Nelson, H.Hirayama and D.W.O. 
Rogers, report number SLAC-265.  

\bibitem{LPM} L.D. Landau and I.J. Pomeranchuk, {\it Dokl. Akad. Nauk. SSSR} 
{\bf 92} (1953) 535 and {\bf 92} (1953) 735. These papers are available in 
English in L. Landau, ``The Collected Papers of L.D. Landau'', Pergamon Press 1965.\\
A.B. Migdal, {\it Phys. Rev.} {\bf 103} (1956) 1811. 

\bibitem{PDG} Particle data table, Phys. Lett. 592 (2004) 1.

\bibitem{Rudi} R. M. Sternheimer, S.M. Seltzer and M.J.Berger, 
Atomic Data and Nuclear Data Tables 30 (1984) 261. 


\bibitem{CORSIKA_PHYSICS.ps.gz} D. Heck et al., Forschungszentrum 
Karlsruhe GmbH, Karlsruhe, Report number FZKA 6019 (1998). 

\bibitem{QGSJET} N.N. Kalmykov and S. Ostapchenko, Phys. Atom. Nucl. 56 (1993)
346, N.N. Kalmykov et al., Nucl. Phys. Proc. Suppl. 52B (1997) 17. 

\bibitem{AZpaper} J. Alvarez-Muniz and E. Zas, {\it Phys. Lett.} 
{\bf B434} (1998) 396 (astro-ph/9806098). 

\bibitem{Knapp} Influence of Hadronic Interaction Model on the Development 
of EAS in Monte Carlo Simulations, D. Heck, J.Knapp and G. Schatz, Nucl. Phys. 
B (Proc. Suppl.) 52B (1997) 139-141. 

\bibitem{Renton} ``An Introduction to the Physics of Quarks and Leptons'' 
by P. Renton (published by Cambridge University Press, 1990)

\bibitem{mks} J. Kwiecinski, A.D. Martin and A.M. Stasto 
Acta Phys. Polon. B31 (2000) 1273 (hep-ph/0004109). 

\bibitem{GYY} A.Z. Gazizov and S.I. Yanush Phys. Rev. D65 (2002) 093003 
(hep-ph/0105368)

\bibitem{Ghandi} R. Ghandi, C. Quigg, M.H. Reno, I. Sarcevic, 
{\it Astroparticle Physics} {\bf 5} (1996) 81. 

\bibitem{MRS99} A.D. Martin, R.G. Roberts, W.J.Stirling and R.S. Thorne,
Eur. Phys. J. C14 (2000) 133 (hep-ph/9907231).

\bibitem{cteq} http://www.phys.psu.edu/~cteq/

\bibitem{KMS1}J. Kwiecinski, A.D. Martin and A.M. Stasto, Phys. Rev. 
D59 (1999) 093002.

\bibitem{KMS2} A.D. Martin, M.G.Ryskin and A.M. Stasto, Acta Phys. Polon. 
B34 (2003)3273. 

\bibitem{RST} R.S. Thorne, private communication.  


\bibitem{Ofelia} O. Pisanti, private communication, see also 
M. Ambrosio et al., astro-ph/0302062.

\bibitem{Herwig} HERWIG, G. Corcella et al., hep-ph/0011363.  

\bibitem{Pierre} ``Introduction to Ultra High Energy Cosmic Rays'' 
by P. Sokolsky (published by Addison-Wesley, 1989).

\bibitem{minuit} F. James and M. Roos, ``Minuit, A System for Function 
Minimization and Analysis of the Parameter Errors and Correlations'',
Comput. Phys. Commun. 10 (1975) 343.

\bibitem{SAUND} J. Vandenbroucke, G. Gratta, N. Lehtenin, 
{\it Astrophys. J.} {\bf 621} (2005) 301. (astro-ph/0406105). 

\bibitem{Justin} J. Vandenbroucke, private communication.

\bibitem{Lehetal} N.G. Lehtinen, S. Adam, G.Gratta, T.K. Berger and M.J. 
Buckingham (astro-ph/0104033).

\bibitem{AM} M.A. Ainslie and J.G. McColm, J.Acoust. Soc. Am 103 (1998) 1671.
   






\end{thebibliography}
\end{document}